\documentclass[prd,final,aps,showpacs]{revtex4}
\usepackage{epsf}
\usepackage{t1enc}
\usepackage[english]{babel}
\usepackage[utf8]{inputenc}
\usepackage{graphicx}
\usepackage{psfrag}
\usepackage{amsmath}
\usepackage{amssymb}
\usepackage{multirow}

\begin{document}

\title{Gravitational waves from binaries on unbound orbits}
\author{J\'anos Maj\'ar$^{1}$, P\'eter Forg\'acs$^{1,2}$, and M\'aty\'as Vas\'uth$^{1}$}
\affiliation{$^1$MTA KFKI Research Institute for Particle and
Nuclear Physics, Budapest 114, P.O.Box 49, H-1525 Hungary\\
$^2$LMPT, CNRS-UMR 6083, Universit\'e de Tours, Parc de Grandmont,
37200 Tours, France}
\date{\today}

\begin{abstract}
A generalized true anomaly-type parametrization, convenient to
describe both bound and open orbits of a two-body system in general
relativity is introduced. A complete description of the time
evolution of both the radial and of the angular equations of a
binary system taking into account the first order post-newtonian
(1PN) is given. The gravitational radiation field emitted by the
system is computed in the 1PN approximation including higher
multipole moments beyond the standard quadrupole term. The
gravitational waveforms in the time domain are explicitly given up
to the 1PN order for unbound orbits, but the results are also
illustrated on binaries on elliptic orbits with special attention
given to the effects of eccentricity.

\end{abstract}

%\pacs{04.25.Nx, 04.30.Db, 97.80.-d}
\maketitle

\section{Introduction}

Among the significant sources of gravitational waves (GWs) for ground-based
interferometers \cite{LIGO,VIRGO,GEO,TAMA} are parabolic and hyperbolic
encounters of compact objects. Similarly to eccentric binaries these sources
have an analytic description which allows one to evaluate the emitted
gravitational wave signal in the post-Newtonian (PN) approximation \cite%
{BlanchetLRR}. During the close encounter of black holes sufficient
amount of energy can be radiated in GWs resulting in the formation
of a binary system. These phenomena are analyzed in
\cite{WalkerWill} with the method of perturbation of orbital
elements. Close encounter type sources provide more characteristic
wave signals than typical elliptic binaries, since after the first
quasiparabolic burstlike signal it emits further bursts
quasiperiodically, until it reaches the validity region of the
elliptic description. Recent estimates show that these binary
interactions can contribute significantly to the sources of
forthcoming advanced gravitational wave observatories \cite{KB}.

A new generation of ground-based interferometric detectors and the
Laser Interferometer Space Antenna (LISA) \cite{Danzmann,BC}
increase the possibility of direct detection of GWs. The LISA
project aims to install a detector system into space and extend the
GW observations to the low-frequency range which is inaccessible to
ground-based interferometers due to the local gravitational noise.
Depending on the system parameters and initial conditions parabolic
and hyperbolic encounters produce GWs in the sensitivity range of
LISA, between $10^{-4}$ and $10^{-1}$ Hz, as it is indicated in our
study of specific binary systems. There are two well-known sources,
the Hulse-Taylor pulsar \cite{HT} and the first double pulsar system
J0737-3039 \cite{Burgay,Willems} representing unique laboratories
for relativistic gravitational physics which provide GWs in the
frequency range of LISA. The computations presented here can be
applied to the advanced ground-based detectors, too.

The classical motion and the encounter of compact objects are
usually considered in the weak field, post-Newtonian regime, where
the motion of the binary system is well described by perturbed
Keplerian orbits. The equations of motion\ and the emitted radiation
of these systems are analyzed in detail in the literature. First
order (1PN) contributions to the dynamical evolution for general,
eccentric orbits are discussed in \cite{DD}. Moreover, the solution
for the radial and angular motion in quasi-Newtonian parametric form
for closed and open orbits was also presented. The contributions of
2PN effects to the periastron advance and the orbital period are
given in \cite{DSch}, where spin-orbit contribution to the secular
precession of the orbit is also analyzed. Spin and tail effects on
the energy and angular momentum losses were determined in \cite{RS}.
The energy and angular momentum fluxes, moreover the evolution of
the orbital elements and the emitted gravitational waveform of
compact binaries to 2PN order are described by \cite{GopIyer} for
eccentric orbits. Their results are expressed by the generalized
quasi-Keplerian description and the eccentric anomaly
parametrization for general elliptic binary orbits. Later, the PN
description of binaries was extended to higher orders and the energy
and angular momentum fluxes are given for inspiralling compact
binaries in quasielliptical orbits at 3PN \cite{ABIQ,ABIS}.
Moreover, 3.5PN corrections to the acceleration for a spherically
symmetric nonrotating self-gravitating star can be found,
\textit{e.g.} in \cite{Itoh}. The 1PN accurate frequency domain
waveforms are given recently in terms of tensor spherical harmonics
for eccentric nonspinning compact binaries in \cite{TessmerSch}.

The main expressions for the Keplerian motion of binaries on unbound orbits
have been derived by many authors in a variety of different contexts. The
energy and angular momentum losses for hyperbolic encounters and the capture
of particles arriving from infinity were analyzed by Hansen \cite{Hansen},
whose work is an extension of the classical results of Peters and Mathews
\cite{PM} for elliptic orbits. In \cite{Turner} the gravitational waveforms
and the energy spectrum from two point masses in arbitrary unbound orbits
were given in the quadrupole formalism. As a result, the waveform is found
to be highly peaked near the periastron for large eccentricities.
Gravitational bremsstrahlung radiation has been studied in \cite{KT} by
considering stars interacting on unbound orbits for the case of large impact
parameters in the postlinear formalism. This description is valid for
arbitrary stellar velocities and mass ratio of the two stars. Recently, the
strain amplitude and the radiation luminosity have been analyzed in \cite%
{SLPIN} for binaries where the periastron distance is much larger than the
Schwarzschild radius of the individual stars.

There are detailed studies of the dynamics of black holes in a stellar
cluster using numerical simulations that include the effects of
gravitational radiation. Cross sections for mergers resulting from
gravitational radiation during two and three-body encounters for a range of
binary semimajor axes and mass ratios were presented in \cite{GMH} with the
use of numerical techniques. In \cite{Martel} fluxes of energy and angular
momentum and the gravitational waveforms produced by a point particle
traveling around a Schwarzschild black hole on arbitrary bound and unbound
orbits are given.

In our work we consider binary systems of compact objects in which
the components are assumed to have negligible rotation. Both the
classical motion (i.e.\ ignoring radiation reaction) and the
radiation field itself are described in the 1PN approximation
including higher multipole moments beyond the quadrupole term. We
introduce a generalized true anomaly parametrization to describe the
radial motion and to express all relevant dynamical quantities in
terms of it. This new definition of the radial parametrization makes
a unified description of both open and closed orbits possible, i.e.\
the parallel discussion of the elliptic, parabolic and hyperbolic
cases. Moreover, we integrate the time dependence of the generalized
true anomaly parameter taking into account the relevant PN
contributions, yielding a complete description of the classical
motion in time of the binary. Our main results are the explicit
solution of the equations of motion on the one hand and the
time-domain gravitational waveforms written out in detail on the
other. To facilitate the use of our expressions, the coefficients of
the waveforms are given in tabular form. We also present detectable
waveforms for realistic sources on eccentric orbits, by numerically
integrating the time dependence of the polarization states. As an
extension of our previous work \cite{MV3} we investigate the emitted
gravitational wave signal of close hyperbolic encounters of compact
objects in detail. Our main results represent the 1PN corrections to
the expressions of Turner \cite{Turner}. These waveforms have a
burstlike character and can be applied to improve present search
algorithms for the detection of gravitational waves.

Our paper is organized as follows. In Sec.\ II we describe the
radial motion of the binary and introduce a suitable radial
parametrization, the generalized true anomaly, which is valid for
all type of orbital motion, namely for the elliptic, parabolic and
hyperbolic cases. We analyze the connection between the coordinate
time and the parametrization in Sec.\
III. In Sec.\ IV we describe the evolution of the polar angle, ${\Upsilon }$%
, in the orbital plane. In Sec.\ V our results are specialized for
the case of circular orbits. Section VI contains our main results,
where the polarization states of the emitted gravitational waves are
given for circular, elliptic and open orbits. In Sec.\ VII
time-dependent gravitational waveforms are computed for such
realistic systems as the Hulse-Taylor and the J0737-3039 pulsars, as
well as binaries on parabolic and hyperbolic orbits, composed of
$8$M$_{\odot}$ and $13$M$_{\odot}$ masses. Section VIII contains our
conclusions and some technical details are relegated to Appendix A.
Appendix B contains the constant coefficients of the waveforms.

Throughout the paper we use units in which $G=c=1$.

\section{Description of the motion and the parametrization of the orbit}

Similarly to \cite{MV3} we introduce two coordinate systems for the
description of the classical motion of the binary and the polarization
states. The $z$ axis of the invariant coordinate system, in which the
evolution of the orbital elements and the dynamics of the binary are most
conveniently described, is fixed to the direction of the total angular
momentum $\mathbf{L}$. The calculations can be simplified if the $x$ and $y$
axes of this system are chosen in a way that the direction of the line of
sight has the components $\mathbf{N}=\left( \sin {\gamma },0,\cos {\gamma }%
\right) $ with the constant angle $\gamma $ of $\mathbf{L}$ and $\mathbf{N}$%
. In the invariant system the separation vector $\mathbf{r}=r\mathbf{n}$ is
written as $\mathbf{r}=r\left( \cos {\Upsilon },\sin {\Upsilon },0\right) $,
where ${\Upsilon }$ is the polar angle of the motion in the orbital plane.

The expressions for the polarization states become simpler with the
description of the main vector quantities and the transverse-traceless (TT) part $%
h_{TT}^{ij}$ of the radiation field in the comoving system. In this
coordinate system the $z$ and $x$ axes are aligned with the
direction of the Newtonian angular momentum $\mathbf{L}_{N}=\mu \mathbf{r}\times \mathbf{v%
}$ and the separation vector $\mathbf{r}$, respectively, where $\mu
=m_{1}m_{2}/m$ is the reduced mass, $m=m_{1}+m_{2}$ is the total mass of the
binary and $\mathbf{v}$ is the relative velocity vector. The transformation
between the two systems is given by a rotation around the $z$ axis with the
angle ${\Upsilon }$.

The true anomaly parametrization, introduced for the description of
Keplerian orbits, can be generalized to the perturbed Kepler
problem. The form of the parametrization is determined by the radial
equation of the motion,
\begin{eqnarray}
\dot{r}^{2} &=&\frac{2E}{\mu }+\frac{2m}{r}-\frac{L^{2}}{\mu ^{2}r^{2}}%
+\left( \dot{r}^{2}\right) _{PN}\ ,  \notag \\
\left( \dot{r}^{2}\right) _{PN} &=&3(3\eta -1)\frac{E^{2}}{\mu ^{2}}+2(7\eta
-6)\frac{Em}{\mu r}-2(3\eta -1)\frac{EL^{2}}{\mu ^{3}r^{2}}+(5\eta -10)\frac{%
m^{2}}{r^{2}}+(-3\eta +8)\frac{mL^{2}}{\mu ^{2}r^{3}}\ ,  \label{dr}
\end{eqnarray}%
where $\eta =\mu /m$. The total energy $E$ and the magnitude $L$ of the
total angular momentum are constants of motion which follow from a
Lagrangian description \cite{KMG}. The turning points, defined by $\dot{r}%
^{2}=0$, are solved up to the required order,
\begin{eqnarray}
r_{_{min}^{max}} &=&\frac{m\mu \pm A_{0}}{-2E}+\delta r_{_{min}^{max}}^{PN}\
,  \notag \\
\delta r_{_{min}^{max}}^{PN} &=&(\eta -7)\frac{m}{4}\pm (\eta +9)\frac{%
m^{2}\mu }{8A_{0}}\mp (3\eta -1)\frac{A_{0}}{8\mu }\ ,  \label{rmaxmin}
\end{eqnarray}%
where $A_{0}=(m^{2}\mu ^{2}+2EL^{2}/\mu )^{1/2}$ is the magnitude of
the Laplace-Runge-Lenz vector. To introduce the radial
parametrization which is valid for all the possible values of the
total energy, first we look at the
following integral:%
\begin{equation*}
\tilde{\Upsilon}(r^{\prime })=\Upsilon (r)|_{r_{min}}^{r^{\prime
}}=\int_{r_{min}}^{r^{\prime }}\frac{\dot{\Upsilon}dr}{\dot{r}}\ .
\end{equation*}%
The equation for the polar angle $\Upsilon $ can be evaluated following\ the
description in \cite{MV3},
\begin{equation}
\dot{\Upsilon}=\frac{L}{\mu r^{2}}\left\{ 1-\left[ (1-3\eta )\frac{E}{\mu }%
+(4-2\eta )\frac{m}{r}\right] \right\} \ .
\end{equation}

As a consequence of the definition $\tilde{\Upsilon}(r_{min})=0$. In the
case of elliptic orbits $r_{min}$ and $r_{max}$ are the turning points of
the radial motion, while for unbounded orbits $r_{min}$ is the only positive
root, and $r_{max}=\infty $. According to these cases the value of $\tilde{%
\Upsilon}(r_{max})$ is the following. For elliptic and parabolic orbits we
have%
\begin{equation*}
\tilde{\Upsilon}(r_{max})|_{elliptic,parabolic}=\pi \mathcal{K}\ ,
\end{equation*}%
where $\mathcal{K}=\left( 1+3m^{2}\mu ^{2}/L^{2}\right) $. In the hyperbolic
case%
\begin{equation*}
\tilde{\Upsilon}(r_{max})|_{hyperbolic}=\chi _{hyp}\mathcal{K}-\frac{m\mu
\sqrt{2E\mu }(\eta L^{2}E-12\mu ^{3}m^{2}-15LE^{2})}{4LA_{0}^{2}}\ ,
\end{equation*}%
where $\chi _{hyp}=\arccos {\left( -m\mu /A_{0}\right) }$ is the maximum
value of $\chi $ for the hyperbolic motion in the Newtonian case.

To get equivalent parametrization with the results of Damour and
Deruelle \cite{DD} and that of \cite{KMG} our generalized true
anomaly parameter$\
\chi $ is introduced as follows%
\begin{equation}
\frac{dr}{d(\cos {\chi })}=-\gamma r^{2}\ ,\qquad \chi _{_{min}^{max}}=%
\tilde{\Upsilon}(r_{_{min}^{max}})/\mathcal{K}.
\end{equation}%
The solution of the radial motion is%
\begin{equation}
r(\chi )=r_{N}(\chi )+r_{PN}(\chi ),  \label{rchi}
\end{equation}%
where
\begin{eqnarray}
r_{N}(\chi ) &=&\frac{L^{2}}{\mu (\mu m+A_{0}\cos {\chi })}  \notag \\
r_{PN}(\chi ) &=&-\frac{2(6-\eta )m^{4}\mu ^{6}+2(10-3\eta )EL^{2}m^{2}\mu
^{3}+(1-3\eta )E^{2}L^{4}}{2A_{0}\mu ^{3}(\mu m+A_{0}\cos {\chi })^{2}}\cos {%
\chi }-\frac{2(2-\eta )mEL^{2}+(6-\eta )m^{3}\mu ^{3}}{\mu (\mu m+A_{0}\cos {%
\chi })^{2}}\ .  \label{rchipert}
\end{eqnarray}%
We note that the above solution is valid for all three types of
orbits.

The relation between the generalized true anomaly parameter and the
coordinate time is given by%
\begin{equation}
\frac{dt}{d\chi }=\frac{1}{\dot{r}}\frac{dr}{d\chi }=\frac{\mu r^{2}}{L}%
\left\{ 1-\frac{1}{2L^{2}}\left[ (\eta -13)m^{2}\mu ^{2}+(3\eta
-1)A_{0}^{2}+(3\eta -8)m\mu A_{0}\cos {\chi }\right] \right\} \ .
\label{tchi}
\end{equation}

The type of orbital motion, namely elliptic, parabolic or
hyperbolic, is determined by $A_{0}$ in Eq.(\ref{rchi}) through the
value of the total energy $E$. For elliptic orbits the energy is
negative and $m\mu >A_{0}$. In this case the length of the
separation vector is changing periodically between the turning
points and the coordinate time $t$ is a monotonically increasing
function of the true anomaly parameter. For parabolic orbits the
energy is zero, $m\mu =A_{0}$, and $r(\chi )$ diverges at $\chi =\pi
$. In this limit $t$ also diverges since $dt/d\chi \sim $ $r^{2}$.
For hyperbolic orbits the energy is negative and $m\mu <A_{0}$. The
divergence of $r(\chi )$ as the parameter approaches the value $\chi
=\cos ^{-1}(-m\mu /A_{0})$ defines the boundary of the orbit.

Because of the fact that the formal expressions of $r(\chi )$ and
$dt/d\chi $ are identical for all types of orbits the integration of
the angular variables formally leads to the same expressions. The
difference between the
three cases arises in the evaluation of the time dependence of $\chi $ and $%
r $, which is determined by the value of $A_{0}$.

\section{The coordinate time and the parameter}

To obtain time-domain gravitational waveforms it is necessary to
express the time dependence of the true anomaly parameter $\chi $.
This time dependence, however, cannot be given analytically. The
most straightforward way is the numerical integration of the
differential equations, nevertheless, in the case of elliptic orbits
the separation of different contributions belonging to different PN
orders is highly nontrivial. As a solution of this problem first we
analyze the analytic form of the inverse function $t(\chi )$ and
determine a general form of the inverse function $\chi (t)$.\

\subsection{Integration of $t(\protect\chi)$}

In the leading Newtonian order the connection between the coordinate time
and the parameter is governed by the relation%
\begin{equation*}
\frac{dt}{d\chi }=\frac{\mu r_{N}^{2}}{L}=\frac{L^{3}}{\mu (m\mu +A_{0}\cos
\chi )^{2}}\ .
\end{equation*}%
The solution of this equation for elliptic orbits is
\begin{equation}
t=\frac{2m\mu ^{3/2}}{(-2E)^{3/2}}A_{ell}+\frac{LA_{0}\sin {\chi }}{%
2E(A_{0}\cos {\chi }+m\mu )}\ ,
\end{equation}%
where
\begin{equation*}
A_{ell}=\mathop{\text{arctg}}\frac{(m\mu -A_{0})\mathop{\text{tg}}\left(
\frac{\chi }{2}\right) }{\sqrt{m^{2}\mu ^{2}-A_{0}^{2}}}\ .
\end{equation*}%
For parabolic orbits we have%
\begin{equation*}
t=\frac{L^{3}\mathop{\text{tg}}{\left( \frac{\chi }{2}\right) }\left[ %
\mathop{\text{tg}}^{2}{\left( \frac{\chi }{2}\right) }+3\right] }{6\mu
^{3}m^{2}}\ ,
\end{equation*}%
while for hyperbolic orbits the result is%
\begin{equation*}
t=-\frac{2m\mu ^{3/2}}{(2E)^{3/2}}A_{hyp}+\frac{LA_{0}\sin {\chi }}{%
2E(A_{0}\cos {\chi }+m\mu )}\ ,
\end{equation*}%
where
\begin{equation*}
A_{hyp}=\mathop{\text{arth}}\frac{(m\mu -A_{0})\mathop{\text{tg}}\left(
\frac{\chi }{2}\right) }{\sqrt{A_{0}^{2}-m^{2}\mu ^{2}}}\ .
\end{equation*}

For a complete description the perturbative solution of the relation between
the coordinate time and $\chi $ is required. Up to 1PN the corrections of $%
r(\chi )$ in Eq.(\ref{tchi}) are taken into account with the
neglection of higher order terms. The solution for elliptic orbits
is
\begin{eqnarray}
t &=&\frac{LA_{0}\sin {\chi }}{2E(A_{0}\cos {\chi }+m\mu )}-\frac{m\mu ^{1/2}%
}{4E(-2E)^{1/2}}\left[ 4\mu +E(\eta -15)\right] A_{ell}  \notag \\
&-&\frac{A_{0}\cos {\chi }[4(3\eta -1)EL^{2}+7(1+\eta )m^{2}\mu ^{3}]+m\mu
\lbrack 4(2\eta +26)EL^{2}+7(31+3\eta )m^{2}\mu ^{3}]}{8A_{0}\mu ^{2}(m\mu
+A_{0}\cos {\chi })^{2}}\ .
\end{eqnarray}%
For parabolic orbits we have%
\begin{equation*}
t=\frac{\left[ (3\eta -18)m^{2}\mu ^{2}+2L^{2}\right] \mathop{\text{tg}}^{2}{%
\left( \frac{\chi }{2}\right) }-\left[ (9\eta +6)m^{2}\mu ^{2}-6L^{2}\right]
}{12\mu ^{3}m^{2}}L\mathop{\text{tg}}\left( \frac{\chi }{2}\right) \ ,
\end{equation*}%
and for hyperbolic orbits the solution is%
\begin{eqnarray}
t &=&\frac{LA_{0}\sin {\chi }}{2E(A_{0}\cos {\chi }+m\mu )}-\frac{m\mu ^{1/2}%
}{4E(2E)^{1/2}}\left[ 4\mu +E(\eta -15)\right] A_{hyp}  \notag \\
&-&\frac{A_{0}\cos {\chi }[4(3\eta -1)EL^{2}+7(1+\eta )m^{2}\mu ^{3}]+m\mu
\lbrack 4(2\eta +26)EL^{2}+7(31+3\eta )m^{2}\mu ^{3}]}{8A_{0}\mu ^{2}(m\mu
+A_{0}\cos {\chi })^{2}}\ .
\end{eqnarray}

\subsection{Integration of $\protect\chi(t)$}

The differential equation which determines the time dependence of the
parameter $\chi $ is
\begin{eqnarray}
\frac{d\chi }{dt} &=&\frac{L}{\mu r^{2}}\left\{ 1+\frac{1}{2L^{2}}\left[
(\eta -13)m^{2}\mu ^{2}+(3\eta -1)A_{0}^{2}+(3\eta -8)m\mu A_{0}\cos {\chi }%
\right] \right\}   \notag \\
&\approx &\frac{L}{\mu r_{N}^{2}}-\frac{2Lr_{PN}}{\mu r_{N}^{3}}+\frac{1}{%
2L\mu r_{N}^{2}}\left[ (\eta -13)m^{2}\mu ^{2}+(3\eta -1)A_{0}^{2}+(3\eta
-8)m\mu A_{0}\cos {\chi }\right] \ ,  \label{dchidt}
\end{eqnarray}%
where $r_{N}$ and $r_{PN}$ are given in Eq. (\ref{rchipert}). Since
we look for the solution as $\chi (t)=\chi _{N}(t)+\chi _{PN}(t)$,
the different corrections to the above equation has to be separated.
This process can easily be done in the case of open orbits, where
$\chi (t)$ is a bounded
function at every PN order. In this case the differential equations for $%
\chi $ are
\begin{eqnarray}
\frac{d\chi _{N}}{dt} &=&\left( \frac{d\chi }{dt}\right) \Bigg|_{N}=\frac{%
\mu (\mu m+A_{0}\cos {\chi _{N}})^{2}}{L^{3}}  \notag \\
\frac{d\chi _{PN}}{dt} &=&\left( \frac{d\chi }{dt}\right) \Bigg|_{PN}=-\frac{%
2A_{0}\chi _{PN}\sin {\chi _{N}}\mu (\mu m+A_{0}\cos {\chi _{N}})}{L^{3}}-%
\frac{2\mu ^{2}(\mu m+A_{0}\cos {\chi _{N}})^{3}r_{PN}(\chi _{N})}{L^{5}}
\notag \\
&+&\frac{\mu (\mu m+A_{0}\cos {\chi _{N}})^{2}}{2L^{5}}\left[ (\eta
-13)m^{2}\mu ^{2}+(3\eta -1)A_{0}^{2}+(3\eta -8)m\mu A_{0}\cos {\chi _{N}}%
\right] \ ,
\end{eqnarray}%
where
\begin{equation}
r_{PN}(\chi _{N})=-\frac{2(6-\eta )m^{4}\mu ^{6}+2(10-3\eta )EL^{2}m^{2}\mu
^{3}+(1-3\eta )E^{2}L^{4}}{2A_{0}\mu ^{3}(\mu m+A_{0}\cos {\chi _{N}})^{2}}%
\cos {\chi _{N}}-\frac{2(2-\eta )mEL^{2}+(6-\eta )m^{3}\mu ^{3}}{\mu (\mu
m+A_{0}\cos {\chi _{N}})^{2}}\ .  \notag
\end{equation}%
Since $\chi $ and $\chi _{PN}$ are bounded functions of time the equations
can be integrated numerically without difficulty.

For elliptic orbits because the time dependence of $\chi _{PN}$ is given by
an unbounded function, therefore simple series expansion gives rise to
secular divergences. To deal with this problem one has to separate the
bounded and unbounded terms in $\chi _{N}$ and $\chi _{PN}$.

To find the solution we note that $t(\chi )$ can be written as
\begin{eqnarray}
t(\chi ) &=&T\chi +f(\chi )  \notag \\
&=&t_{N}(\chi )+t_{PN}(\chi )=(T_{N}+T_{PN})\chi +f_{N}(\chi )+f_{PN}(\chi
)\ ,
\end{eqnarray}%
where $T_{N}$ and $T_{PN}$ are constants, and the functions $f_{N}(\chi )$
and $f_{PN}(\chi )$ are bounded. To prepare the numerical inversion of $%
t(\chi )$, first we use the following general theorem about smooth functions.%
\newline
Let $F(x)$ be a smooth function, which can be written in the form $%
y=F(x)=Ax+B(x)$, where $A\neq 0$ is constant, and $B(x)$ is a bounded
function of $x$. If the $F^{-1}$ inverse function exists, writing it in the
form $F^{-1}(y)=y/A+b(y)$, $b(y)$ will be a bounded function of $y$.\newline
With the use of this theorem we look for the solution in the following form:
\begin{eqnarray}
\chi (t) &=&\Omega t+g(t)  \notag \\
&=&\Omega _{N}t+g_{N}(t)+\Omega _{PN}t+g_{PN}(t)\ .
\end{eqnarray}%
The constants $T_{N}$, $T_{PN}$, and hence $\Omega _{N}$ and $\Omega _{PN}$
can be determined by the investigation of the circular orbit limit. The
results are
\begin{equation*}
\Omega _{N}=\frac{(-2E)^{3/2}}{m\mu ^{3/2}}\ ,\qquad \Omega _{PN}=-\frac{%
(\eta -15)E^{3}}{(-2E)^{1/2}m\mu ^{5/2}}\ .
\end{equation*}

Moreover, we note that the use of the straightforward series
expansion during the calculation of the sine and cosine of $\chi
(t)$ results in the appearance of secular divergences. To solve this
problem, as an example, the following expression is to be used:
\begin{equation*}
\sin {\chi (t)}=\sin {[\Omega _{N}t+\Omega _{PN}t+g_{N}(t)]}+\cos {[\Omega
_{N}t+\Omega _{PN}t+g_{N}(t)]}g_{PN}(t)\ ,
\end{equation*}%
and similarly for $\cos {\chi (t)}$, which leads to a combined amplitude and
frequency series expansion. In our calculations we use the above expansion
to avoid secular divergences in the description of the motion of the binary
and the time dependence of the polarization states of the detectable
waveform.

Although the above expansion helps us to avoid the appearance of
secular terms, it gives rise to technical difficulties when
separating the different PN corrections on the level of
Eq.(\ref{dchidt}). With the integration of both the Newtonian order
and the full equation, one can determine $g_{N}(t)$ and $g_{PN}(t)$
separately, and give the full description of the time dependence of
$\chi $. The details of the calculations are given in Appendix A.

\section{Angular evolution}

Similarly to $r$ the polar angle $\Upsilon $ is decomposed as

\begin{equation*}
\Upsilon =\Upsilon _{N}+\Upsilon _{PN}\ .
\end{equation*}%
To solve the equations of motion the form of the relative velocity
vector is required in terms of the Euler angles. In the invariant
system it is expressed as $\mathbf{v}=\left(
\dot{r},r\dot{\Upsilon},0\right) $. For the description of
gravitational waves and the equations of motion one needs an
alternative form of the components of the velocity vector, both in
their general form and in terms of the parametrization.

In terms of the constants of motion one can express $\dot{r}^{2}$, Eq.(\ref%
{dr}), and the square of the velocity vector, which is%
\begin{equation*}
v^{2}=\frac{2E}{\mu }+\frac{2m}{r}-\left[ 3(1-3\eta )\left( \frac{E}{\mu }%
\right) ^{2}+2(6-7\eta )\frac{Em}{\mu r}+(10-5\eta )\left( \frac{m}{r}%
\right) ^{2}-\frac{\eta mL^{2}}{\mu ^{2}r^{3}}\right] \ .
\end{equation*}%
From our previous results if follows that $v_{\parallel }^{2}=\dot{r}^{2}$
and, moreover,
\begin{equation}
v_{\perp }^{2}=v^{2}-\dot{r}^{2}=\frac{L^{2}}{\mu ^{2}r^{2}}-(2-6\eta )\frac{%
EL^{2}}{\mu ^{3}r^{2}}-(8-4\eta )\frac{mL^{2}}{\mu ^{2}r^{3}}\ .  \notag
\end{equation}%
This indicates that%
\begin{equation*}
v_{\perp }=\frac{L}{\mu r_{N}}-\frac{L}{\mu r_{N}}\left[ \frac{r_{PN}}{r_{N}}%
+(1-3\eta )\frac{E}{\mu }+(4-2\eta )\frac{m}{r_{N}}\right] \ .
\end{equation*}%
When we insert the parameter dependence of $r_{N}$ and $r_{PN}$ in the above
formulas the components of the velocity vector become%
\begin{equation}
v_{\parallel }=\frac{A_{0}\sin {\chi }}{L}+\frac{2(\eta -1)m^{4}\mu
^{6}+3(3\eta -1)E^{2}L^{4}+2(4\eta -5)EL^{2}m^{2}\mu ^{3}}{2\mu
^{2}A_{0}L^{3}}\sin {\chi }-\frac{(8-3\eta )m\mu A_{0}^{2}}{2L^{3}}\ ,
\end{equation}%
and
\begin{eqnarray}
v_{\perp } &=&\frac{\mu m+A_{0}\cos {\chi }}{L}-\frac{2(9-10\eta )m^{4}\mu
^{6}+2(7-8\eta )EL^{2}\mu ^{3}m^{2}+3(1-3\eta )E^{2}L^{4}}{2\mu
^{2}A_{0}L^{3}}\cos {\chi }+ \\
&&\frac{(2+\eta )m^{3}\mu ^{3}+(3+\eta )mEL^{2}}{2L^{3}}+\frac{2(\eta
-2)m^{3}\mu ^{3}+4(\eta -2)mEL^{2}}{L^{3}}\cos ^{2}{\chi }\ ,
\end{eqnarray}%
From the relation $v_{\perp }=r\dot{\Upsilon}$ we have%
\begin{equation}
\dot{\Upsilon}=\frac{L}{\mu r_{N}^{2}}\left\{ 1-\left[ \frac{2r_{PN}}{r_{N}}%
+(1-3\eta )\frac{E}{\mu }+(4-2\eta )\frac{m}{r_{N}}\right] \right\} \ ,
\label{Upsilondot}
\end{equation}%
which can be expressed in terms of the parametrization as
\begin{equation}
\left( \frac{d\Upsilon }{d\chi }\right) =1-\frac{6m^{2}\mu ^{2}+\eta m\mu
A_{0}\cos {\chi }}{2L^{2}}\ .  \notag
\end{equation}%
The solution for $\Upsilon $ is%
\begin{equation}
\Upsilon =\Upsilon _{0}+\chi -\frac{6m^{2}\mu ^{2}\chi +\eta m\mu A_{0}\sin {%
\chi }}{2L^{2}}\ .  \label{upsi}
\end{equation}%
The separation vector and the waveforms contain expressions of $\sin {%
\Upsilon }$ and $\cos {\Upsilon }$, and their higher harmonics. With the use
of a series expansion, for example, $\sin {\Upsilon }$ can be expressed as
\begin{equation*}
\sin {\Upsilon }=\sin {(\Upsilon _{0}+\chi )}-\cos {(\Upsilon _{0}+\chi )}%
\left( \frac{6m^{2}\mu ^{2}\chi +\eta m\mu A_{0}\sin {\chi }}{2L^{2}}\right)
\ .
\end{equation*}%
For elliptic orbits there is no upper limit for $\chi $ as a function of the
coordinate time and the $\left( 3m^{2}\mu ^{2}/L^{2}\right) \chi \cos {%
(\Upsilon _{0}+\chi )}$\ perturbative term gives rise to a secular
divergence. To handle this problem we follow the method described above in
the case of the time dependence of the parameter, namely, instead of the
above expression we apply the following decomposition:%
\begin{equation*}
\sin {\Upsilon }=\sin {\left( \Upsilon _{0}+\chi -\frac{3m^{2}\mu ^{2}\chi }{%
L^{2}}\right) }-\cos {\left( \Upsilon _{0}+\chi -\frac{3m^{2}\mu ^{2}\chi }{%
L^{2}}\right) }\frac{\eta m\mu A_{0}\sin {\chi }}{2L^{2}}\ ,
\end{equation*}%
where no linear terms appear (note that a similar expression is valid for $%
\cos {\Upsilon )}$.

\section{Equations of motion for circular orbits}

The relative velocity vector can generally be written as $\mathbf{v}=\dot{r}%
\mathbf{n}+r\omega \mathbf{m}$, where $\mathbf{m}$ is the unit vector in the
direction of the $y$ axis of the comoving system. The definition of a
circular orbit is
\begin{equation*}
\dot{r}=\dot{\omega}=0\ .
\end{equation*}%
From the above decomposition of the relative velocity it turns out that $%
\omega =\dot{\Upsilon}.$ As a consequence,

\begin{equation}
v=v_{\perp }=\frac{L}{\mu r}\left[ 1-(1-3\eta )\frac{E}{\mu }-(4-2\eta )%
\frac{m}{r}\right] \ .  \label{vcirc}
\end{equation}%
In the circular orbit limit the constant radius $r$ of the orbit
cannot be evaluated analytically in all orders of the approximation
and hence in our calculations we keep $r$ as a constant parameter
determined later.

From Eq.(\ref{vcirc}) the time evolution of $\Upsilon $ can be written as%
\begin{equation}
\Upsilon =\frac{Lt}{\mu r^{2}}\left[ 1-(1-3\eta )\frac{E}{\mu }-(4-2\eta )%
\frac{m}{r}\right] \ .
\end{equation}

During the evaluation of quantities including the sine and cosine of $%
\Upsilon $ there are difficulties occurring in connection with the
secular divergences as in the elliptic case. The solution of this
problem is the same, although it leads to simpler form. For example,
\begin{equation}
\sin {\Upsilon }=\sin {\left\{ \frac{Lt}{\mu r^{2}}\left[ 1-(1-3\eta )\frac{E%
}{\mu }-(4-2\eta )\frac{m}{r}\right] \right\} }\ .  \notag
\end{equation}%
In this case the secular terms can be treated with a simple expansion series
of the orbital frequency.

\section{The structure of the detectable waveform}

The parameter and, for circular orbits, time dependence of the waveform and
the polarization states $h_{+}$ and $h_{\times }$ are determined by the use
of the method described in \cite{MV3}. The polarization states of the
detectable gravitational waves are calculated by the projections of the
transverse-traceless tensor $h_{TT}^{ij}$ representing metric perturbations,
\begin{equation*}
h_{+}=\frac{1}{2}(p_{i}p_{j}-q_{i}q_{j})h_{TT}^{ij}\ ,\quad h_{\times }=%
\frac{1}{2}(p_{i}q_{j}+q_{i}p_{j})h_{TT}^{ij}\ ,
\end{equation*}%
where $\mathbf{p}$ is a unit vector in the orbital plane perpendicular to
the direction of the line of sight $\mathbf{N}$, and $\mathbf{q}=\mathbf{N}%
\times \mathbf{p}$.

Since we have chosen the comoving system to describe this projection of $%
h_{TT}^{ij}$ we determine the components of $\mathbf{N}$, $\mathbf{p}$ and $%
\mathbf{q}$ in this system. With the use of the transformation law
between the two coordinate systems $\mathbf{N}$ has the following
form in the comoving one:
\begin{equation*}
\mathbf{N}=\left(
\begin{array}{c}
\sin {\gamma }\cos {\Upsilon } \\
-\sin {\gamma }\sin {\Upsilon } \\
\cos {\gamma }%
\end{array}%
\right) \ .
\end{equation*}%
Moreover, the conditions for $\mathbf{p}$ determine its components as
\begin{equation*}
\mathbf{p}=\left(
\begin{array}{c}
\sin {\Upsilon } \\
\cos {\Upsilon } \\
0%
\end{array}%
\right) \ .
\end{equation*}%
To separate the different contributions the transverse-traceless part of the
radiation field $h_{TT}^{ij}$ can be decomposed in the post-Newtonian
approximation as \cite{Kidder}
\begin{equation*}
h_{TT}^{ij}=\frac{2\mu }{D}\left[ Q^{ij}+P^{0.5}Q^{ij}+PQ^{ij}\right] _{TT}\
,
\end{equation*}%
where $D$ is the distance between the observer and the source and we have
collected all the terms which are relevant up to 1 PN order. $Q^{ij}$
denotes the quadrupole (or Newtonian) term, $P^{0.5}Q^{ij}$, and $PQ^{ij}$
are higher order relativistic corrections. The detailed expressions for
these contributions are given in \cite{Kidder,WW}.

We choose a similar decomposition for the polarization states $h_{+}$ and $%
h_{\times }$,
\begin{equation*}
h_{_{\times }^{+}}=\frac{2\mu }{D}\left[ h^{N_{\times }^{+}}+h^{0.5_{\times
}^{+}}+h^{1_{\times }^{+}}\right] \ ,
\end{equation*}%
It is worth mentioning that this decomposition is formal in the
sense that higher order terms in the description of motion can cause
higher order contributions in $h^{N_{\times }^{+}}$.

Since the description of the motion of the binary system has
different properties for the different types of the orbit, and hence
one can use different simplifications throughout the evaluations,
the elliptic, circular, and open orbit cases are discussed
separately including the description of the simplifications and the
structure of the detectable waveform. The constant coefficients are
given in detail in Appendix B.

\subsection{The elliptic orbit case}

For elliptic orbits a combined series expansion has to be included
to deal with the problem of secular divergences described above.
These secular terms give rise to a frequencylike series beside the
usual amplitude expansion. In order to define this we introduce
\begin{equation}
\chi ^{\prime }=\Upsilon _{0}+\chi -\frac{3m^{2}\mu ^{2}}{L^{2}}\chi ,
\label{chiprime}
\end{equation}%
which is valid up to 1PN. To lowest, Newtonian order the above
definition becomes simply $\chi ^{\prime }=\Upsilon _{0}+\chi $.
Since our goal is to give waveform expressions valid up to 1PN in
the following in our expressions the original definition of $\chi
^{\prime }$ will be used.

Following the main steps described above, the parameter dependence
of the detectable gravitational waveform for eccentric orbits can be
given as
\begin{equation*}
h^{N_{\times }^{+}}=\sum_{k=-2}^{2}\sum_{j=-3}^{3}\big[C_{k,j}^{N_{\times
}^{+}}\cos {(k\chi ^{\prime }+j\chi )}+S_{k,j}^{N_{\times }^{+}}\sin {(k\chi
^{\prime }+j\chi )}\big]\ ,
\end{equation*}%
\begin{equation*}
h^{0.5_{\times }^{+}}=\sum_{k,j=-3}^{3}\big[C_{k,j}^{0.5_{\times }^{+}}\cos {%
(k\chi ^{\prime }+j\chi )}+S_{k,j}^{0.5_{\times }^{+}}\sin {(k\chi ^{\prime
}+j\chi )}\big]\ ,
\end{equation*}%
\begin{equation}
h^{1_{\times }^{+}}=\sum_{k,j=-4}^{4}\big[C_{k,j}^{1_{\times }^{+}}\cos {%
(k\chi ^{\prime }+j\chi )}+S_{k,j}^{1_{\times }^{+}}\sin {(k\chi
^{\prime }+j\chi )}\big]\ , \label{helliptic}
\end{equation}%
where the $S$ and $C$ coefficients are depending on the energy $E$,
the magnitude of the total angular momentum $L$, the masses of the
objects and the $\gamma $ angle, and hence they are constants. The
explicit form of these constant coefficients is given in Appendix B.

It is worth mentioning that the above expansions are formal, since
the time dependence of the true anomaly parameter $\chi $ contains
higher order corrections. These can modify the expressions for both
the secular terms and the amplitude expansion. Completing the
analysis one arrives to a combined frequency and amplitude series
expansion. In this work the whole method is used to evaluate
(numerically) the time dependence of the polarization states shown
in the next chapter.

\subsection{The circular orbit case}

As it has been shown in the circular orbit case we use the $r$ constant
length of the separation vector as a parameter throughout our calculations.
In this case Eq.(\ref{chiprime}) reduces to
\begin{equation*}
\chi ^{\prime }=\Upsilon _{0}+\frac{Lt}{\mu r^{2}}-\frac{Lt}{\mu ^{2}r^{3}}%
[(1-3\eta )Er+(4-2\eta )m\mu ]=\Upsilon _{0}+\omega _{N}t+\omega _{PN}t\ .
\end{equation*}

With the use of this notation the structure of the detectable waveform
becomes
\begin{equation*}
h^{N_{\times }^{+}}=\sum_{k=0}^{2}\big[C_{k}^{N_{\times }^{+}}\cos {(k\chi
^{\prime })}+S_{k}^{N_{\times }^{+}}\sin {(k\chi ^{\prime })}\big]\ ,
\end{equation*}%
\begin{equation*}
h^{0.5_{\times }^{+}}=\sum_{k=0}^{3}\big[C_{k}^{0.5_{\times }^{+}}\cos {%
(k\chi ^{\prime })}+S_{k}^{0.5_{\times }^{+}}\sin {(k\chi ^{\prime })}\big]\
,
\end{equation*}%
\begin{equation}
h^{1_{\times }^{+}}=\sum_{k=0}^{4}\big[C_{k}^{1_{\times }^{+}}\cos
{(k\chi ^{\prime })}+S_{k}^{1_{\times }^{+}}\sin {(k\chi ^{\prime
})}\big]\ .
\end{equation}%
Again, the $S$ and $C$ constant coefficients are depending on $E$,
$L$, the masses, $r$, and the angle $\gamma $. The explicit form of
these coefficients is given in Appendix B.

The high frequency terms in $h^{N}$ arise from the 1PN terms of the
description of motion. At the Newtonian order the frequency of the
detectable waveform is twice the orbital frequency $\omega _{N}$, in
agreement with the quadrupole formalism.

\subsection{Waveforms for open orbits}

The evaluation process of the waveform is almost the same as it is
in the elliptic case. The only difference is that there are no
secular terms arising in the parabolic and hyperbolic cases, and it
is possible to use the usual expansion of the sine and cosine of the
angular variables. The parameter dependence of the polarization
states can be written as
\begin{equation*}
h^{N_{\times }^{+}}=\chi \sum_{k=0}^{3}\big[C_{k\chi }^{N_{\times }^{+}}\cos
{(k\chi )}+S_{k\chi }^{N_{\times }^{+}}\sin {(k\chi )}\big]+\sum_{k=0}^{4}%
\big[C_{k}^{N_{\times }^{+}}\cos {(k\chi )}+S_{k}^{N_{\times }^{+}}\sin {%
(k\chi )}\big]\ ,
\end{equation*}%
\begin{equation*}
h^{0.5_{\times }^{+}}=\sum_{k=0}^{6}\big[C_{k}^{0.5_{\times }^{+}}\cos {%
(k\chi )}+S_{k}^{0.5_{\times }^{+}}\sin {(k\chi )}\big]\ ,
\end{equation*}%
\begin{equation}
h^{1_{\times }^{+}}=\sum_{k=0}^{7}\big[C_{k}^{1_{\times }^{+}}\cos {(k\chi )}%
+S_{k}^{1_{\times }^{+}}\sin {(k\chi )}\big]\ ,
\end{equation}%
where the constant coefficients are depending on the same parameters as in
the elliptic case, and, moreover, on the $\Upsilon _{0}$ initial value of $%
\Upsilon $. We note that the first type of terms in $h^{N}$ are
linear in the parameter. In these terms all the coefficients are at
1PN order, and since the value of the parameter is bounded, the
description is self-consistent.

\section{Realistic waveform signals}

Following the description above, after the integration of the time
dependence of the generalized true anomaly parameter the realistic,
time-dependent gravitational waveforms can be given. In this section we
present Newtonian waveforms as well as 0.5PN and 1PN corrections for
well-known or reference systems in order to analyze the effects of the
orbital eccentricity.

\subsection{Waveform of realistic elliptic binaries}

For further investigation, we numerically integrate the time
dependence of the polarization states of the detectable waveform for
two realistic and well-known sources, namely, the Hulse-Taylor and
J0737-3039 pulsars, see Table I. Our goal is to investigate the
effects of the eccentricity of the orbit and the 0.5PN and 1PN
corrections to the waveform.

\begin{table}[th]
\begin{center}
\begin{tabular}{|c||c|c|}
\hline
Parameters & Hulse-Taylor & \qquad J0737-3039 \quad \qquad \\ \hline\hline
Mass of object 1 ($m_{1}$) & 1.441 M$_{\odot }$ & 1.337 M$_{\odot }$ \\
Mass of object 2 ($m_{2}$) & 1.387 M$_{\odot }$ & 1.25 M$_{\odot }$ \\
Eccentricity ($\epsilon $) & 0.617 & 0.0878 \\
Periastron ($a$) & 1.9501$\cdot $10$^{9}$ [m] & 0.8787$\cdot
$10$^{9}$ [m]
\\
Distance from detector ($D$) & 21000 [ly] & 1956 [ly] \\
Orbital frequency ($T_{orb}$) & 27907 [s] & 8834 [s] \\
$\mathbf{J}$--$\mathbf{N}$ angle $(\gamma )$ & 21.12$^{\circ }$ & 28.89$%
^{\circ }$ \\ \hline
\end{tabular}%
\end{center}
\caption{Parameters of the investigated sources}
\label{tab:ellparam}
\end{table}

To keep our figures transparent, the $h_{+}$ and $h_{\times }$
polarization states will be shown on the same plot. We collect the
results belonging to different sources and PN orders. This
simplification allows us to show the main differences between the
selected systems order by order.

Following these guidelines, first we show the polarization states at
the Newtonian order.

\begin{figure}[th]
\noindent\hfil
\includegraphics[scale=0.9]{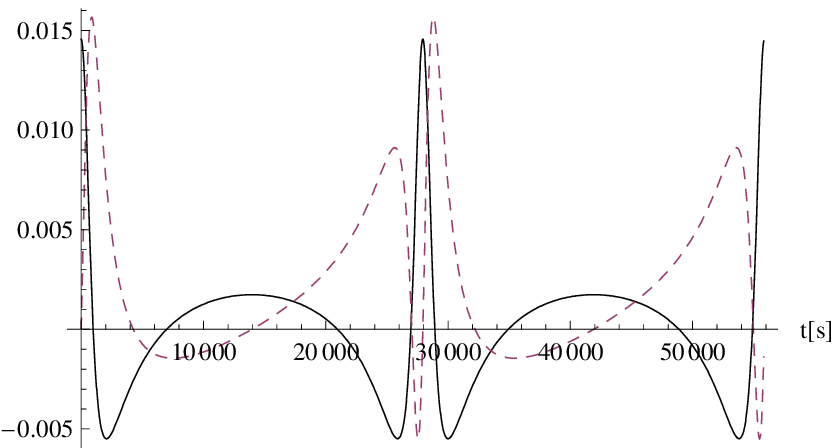}\qquad %
\includegraphics[scale=0.9]{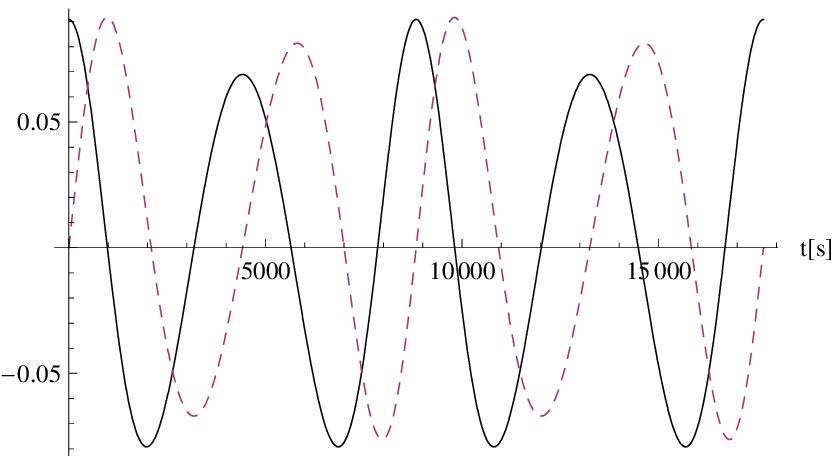}
\caption{Time dependence of the polarization states ($h_{+}$
depicted by solid line and  $h_{\times }$ corresponds to dashed
line) at the Newtonian order for the Hulse-Taylor (left) and for the
J0737-3039 pulsars (right). The amplitudes are multiplied by the
factor $10^{20}$.} \label{fig:HTN}
\end{figure}

Figure \ref{fig:HTN} shows that the eccentricity of the orbit leads
to many differences between the waveform predictions of the
different sources. In the case of the Hulse-Taylor pulsar, the
results for the emitted waveform are significantly different from
the usual circular orbit case, since the eccentricity gives rise to
many higher harmonics of the waveform. In the case of J0737-3039
pulsar, the eccentricity is almost 1 order of magnitude smaller and
the emitted waveform is similar to the one generated by binaries on
the circular orbit. We note that these differences are much more
significant than the higher corrections of the PN series in both
cases.

The amplitude of the emitted waveform of J0737-3039 is much higher than that
of the Hulse-Taylor pulsar since the periastron of the former is less than
half of the latter one, and it is much closer to the detector.

The higher order corrections show the same properties of the
waveforms for the different sources. It is worth mentioning that in
the circular orbit limit the frequency of the emitted waveform is
twice the orbital frequency of the motion.

\begin{figure}[th]
\noindent\hfil
\includegraphics[scale=0.9]{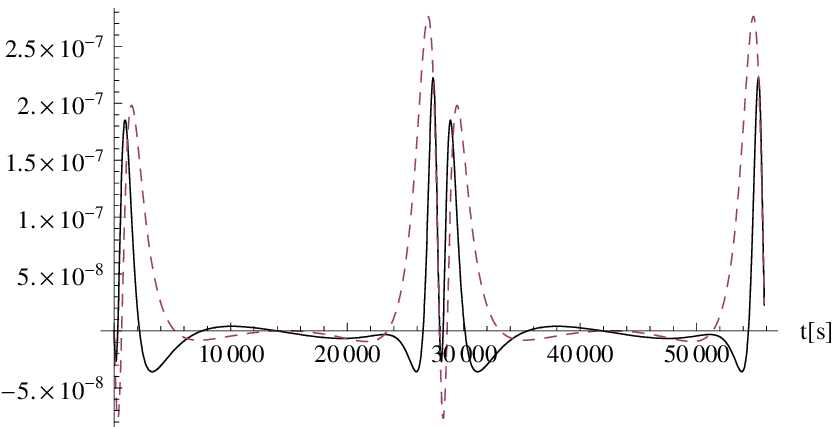}\qquad %
\includegraphics[scale=0.9]{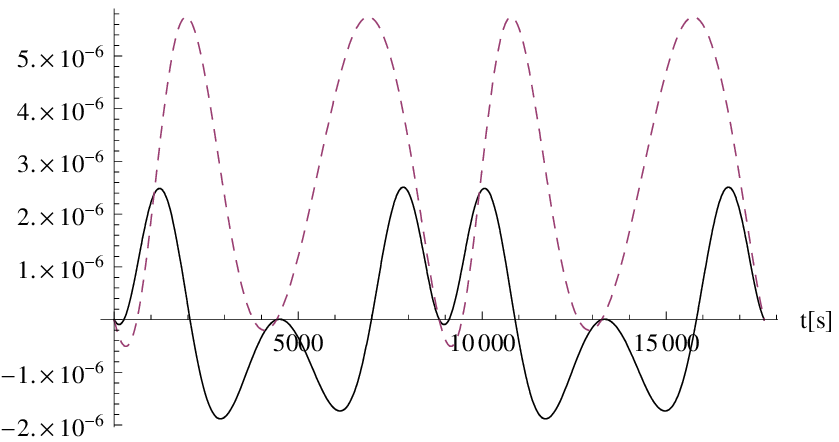}
\caption{Time dependence of the first corrections to the
polarization states at 0.5PN order. Again, $h_{+}$ is depicted by
the solid line and $h_{\times }$ by the dashed line. The corrections
for the Hulse-Taylor pulsar are shown on the left and for the J-0737
pulsar on the right. For further comparison amplitudes are
multiplied by the factor $10^{20}$.} \label{fig:HT05}
\end{figure}

\begin{figure}[th]
\noindent\hfil
\includegraphics[scale=0.9]{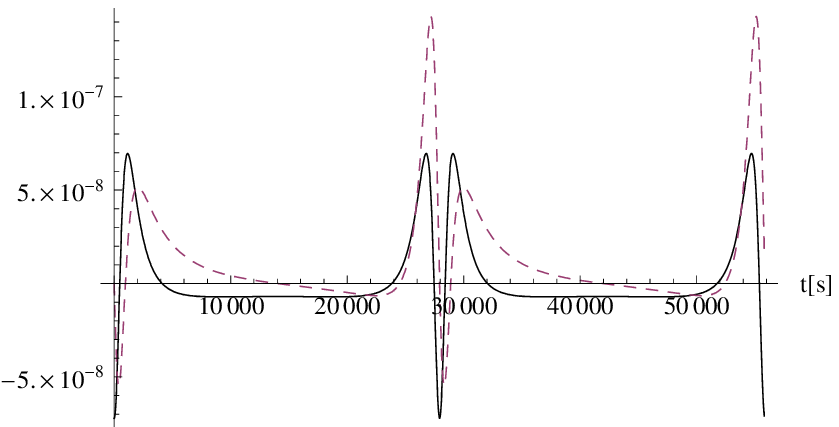}\qquad %
\includegraphics[scale=0.9]{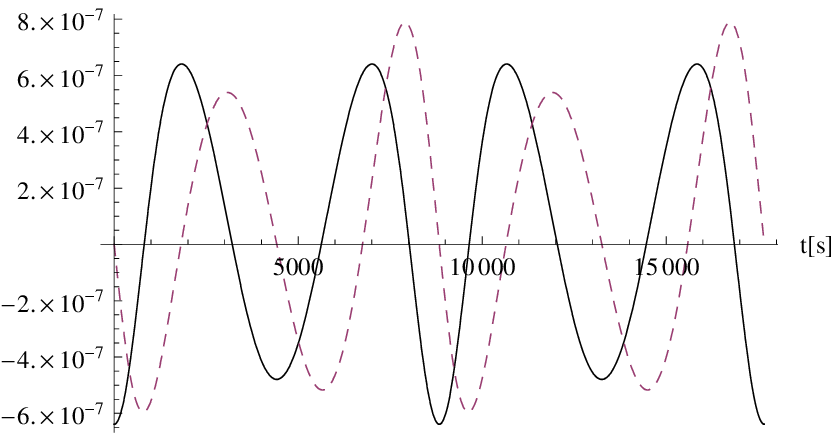}
\caption{Multipolar corrections to the polarization states at 1PN order.
Again, the corrections for the Hulse-Taylor pulsar are shown on the left,
and for the J0737-3039 pulsar on the right. Solid line shows the $h_{+}$
polarization state, and dashed line shows $h_{\times }$. Again, the
amplitudes are multiplied by the factor $10^{20}$.}
\label{fig:HTPN}
\end{figure}

\begin{figure}[th]
\noindent\hfil
\includegraphics[scale=0.9]{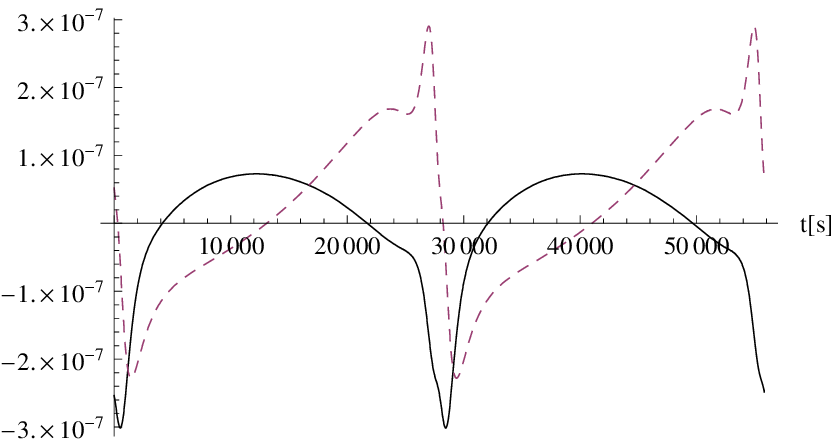}\qquad %
\includegraphics[scale=0.9]{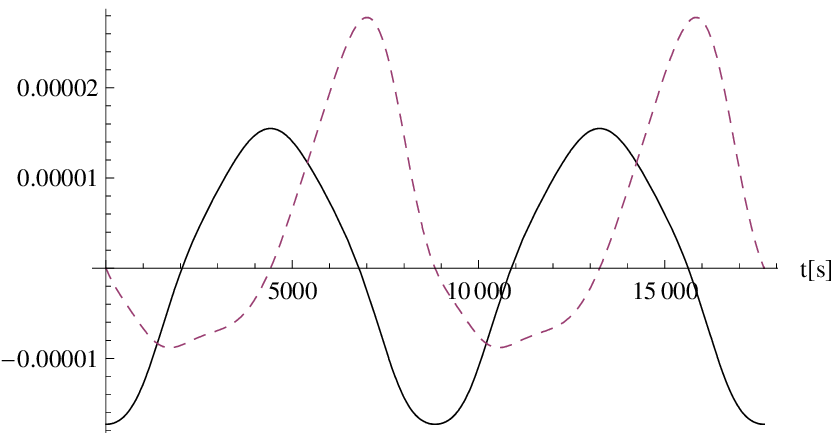}
\caption{1PN order corrections to the polarization states, which
arise from the perturbative description of the elements of the
motion. The corrections for the Hulse-Taylor pulsar are shown on the
left and for the J0737-3039 pulsar on the right. The solid line
shows the $h_{+}$ polarization state and the dashed line shows
$h_{\times }$. To compare the different terms of these corrections,
the amplitudes are multiplied by the same factor as before, namely
$10^{20}$.} \label{fig:HTNPN}
\end{figure}

\begin{figure}[th]
\noindent\hfil
\includegraphics[scale=0.9]{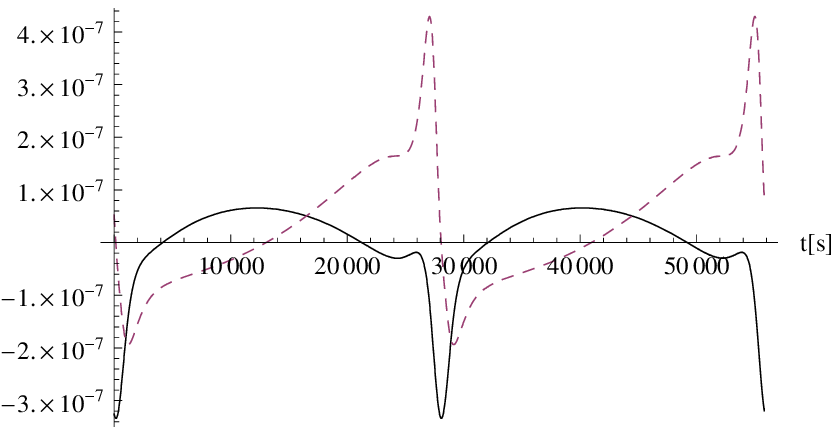}\qquad %
\includegraphics[scale=0.9]{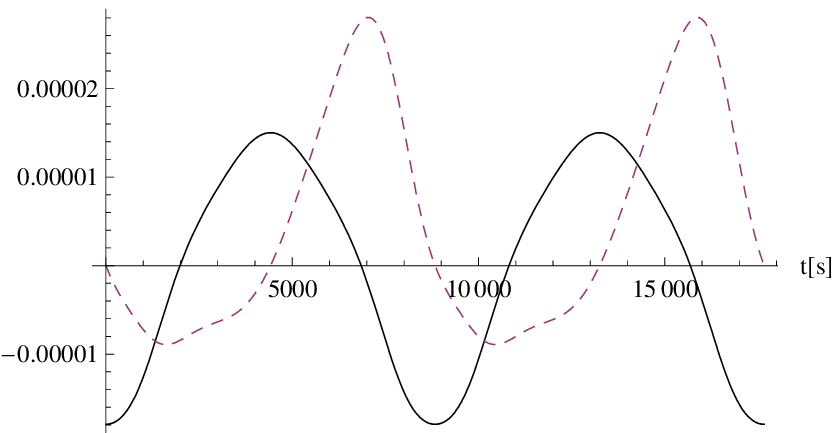}
\caption{In these figures we plot the full 1PN corrections to the
polarization states for the Hulse-Taylor pulsar (left) and for the
J0737-3039 pulsar (right). The solid line represents the $h_{+}$
polarization state and the dashed line shows $h_{\times }$.}
\label{fig:HT1PN}
\end{figure}

Beyond the properties described above there are two additional
effects on the waveform shown by Figs.
\ref{fig:HT05}-\ref{fig:HT1PN}. By comparing the leading amplitudes
of the different corrections it turns out that, since the masses of
the objects are almost equal for both sources, and the 0.5PN order
corrections are multiplied by $\delta m$, these corrections to the
waveform are suppressed by the simple 1PN corrections.

Since the lowest order corrections to the dynamical variables arise
at 1PN, the 0.5PN corrections to the waveform are originated only
from the multipolar expansion. For 1PN corrections both the
multipolar expansion and the contributions from the description of
the motion contribute to the waveform. On the other hand 1PN
corrections from the orbital motion of the binaries are much more
significant than the contributions from the multipolar expansion of
the post-Minkowskian series.

Although a full analysis of the waveform would require Fourier expansion,
the main properties of these realistic waveforms are transparent. The
contributions which can be seen only on long-term calculations are from the
1PN corrections to the frequency of the waveform arising through the time
dependence of true anomaly parameter and the periastron precession. These
effects cause the slow shift of the basic frequency of the waveform.

\subsection{Waveform of reference parabolic and hyperbolic orbits}

Our main goal is to evaluate the first corrections to the detectable
gravitational waves emitted by binaries scattering on hyperbolic and
parabolic orbits.

To investigate the properties of the emitted waveforms of scattering
binary systems, we introduce two hypothetical reference binaries,
see Table II. The masses of the orbiting bodies are chosen following
the parameter estimations in \cite{KB}, although we determined the
values in a way which avoids the suppression of the 0.5PN
corrections. The minimal distance between the objects and the
distance from the detector are chosen to be the same as in the case
of the Hulse-Taylor pulsar. It will give the opportunity to compare
the waveform properties between open and closed orbits.

To investigate sources in which the emission of GWs results in the formation
of a closed binary (in extreme cases even zoom-whirl orbits) we discuss
parabolic sources. To analyze the effects of the eccentricity, besides
parabolic orbits we introduce a hyperbolic source ($\epsilon =2$).

\begin{table}[!ht]
\begin{center}
\begin{tabular}{|c||c|}
\hline
Parameters & \qquad Hypothetic sources \quad\qquad \\ \hline\hline
Mass of object 1 ($m_1$) & 8 M$_\odot$ \\
Mass of object 2 ($m_2$) & 13 M$_\odot$ \\
Eccentricity ($\epsilon$) & 1 and 2 \\
Minimal distance ($r_{min}$) & 1.9501$\cdot$10$^9$ [m] \\
Distance from detector ($D$) & 21000 [ly] \\
$\mathbf{J}$--$\mathbf{N}$ angle $(\gamma)$ & 45$^{\circ}$ \\ \hline
\end{tabular}%
\end{center}
\caption{Parameters of the investigated open sources}
\label{tab:openparam}
\end{table}

\begin{figure}[th]
\noindent\hfil
\includegraphics[scale=0.9]{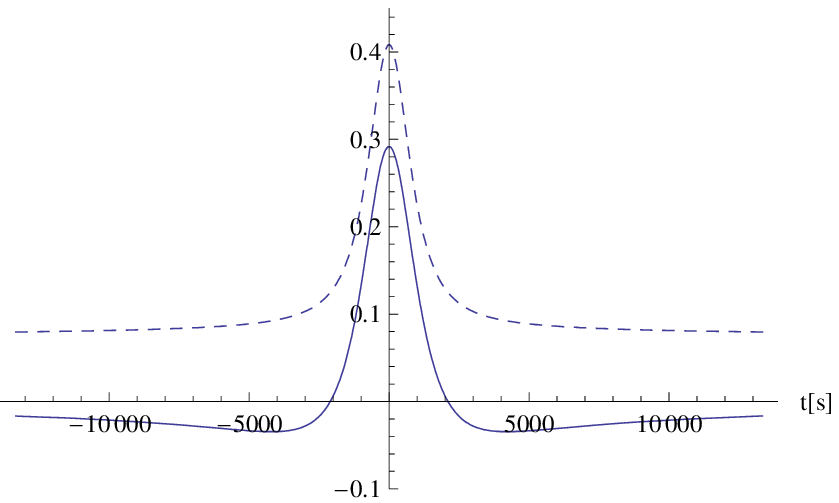}\qquad %
\includegraphics[scale=0.9]{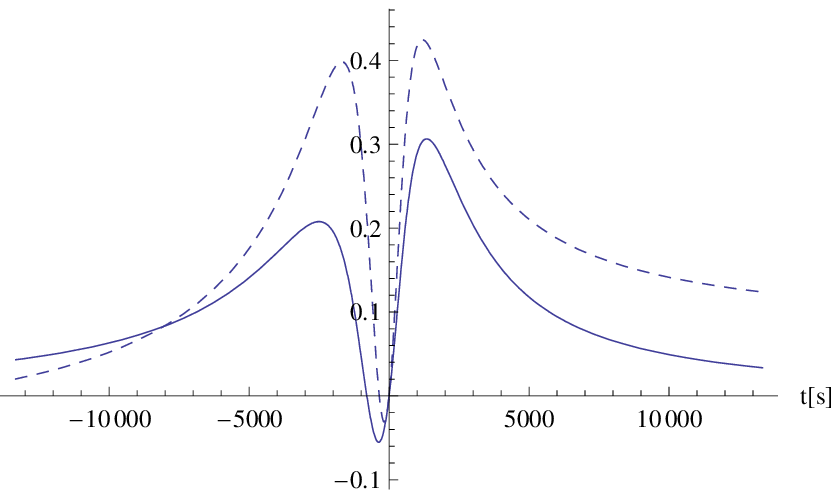}
\caption{Time dependence of the polarization states at Newtonian
order for the two open orbit sources. The solid line shows the
parabolic case and the dashed line shows the waveform of the
hyperbolic ($\protect\epsilon =2$)
source. The $h_{+}$ polarization state are depicted on the left figure and $%
h_{\times }$is shown on the left one. The amplitudes are multiplied by the
factor $10^{20}$.}
\label{fig:openN}
\end{figure}

The waveform for open orbits is not periodic (or quasiperiodic), it
is burstlike, as it is shown in \cite{Turner}. Our hypothetic
sources provide a high amplitude signal in the scale of a few hours.
Binaries in \cite{KB} provide gravitational waves with higher
amplitude, since the minimal distance between the objects are much
smaller than ours.

\begin{figure}[th]
\noindent\hfil
\includegraphics[scale=0.9]{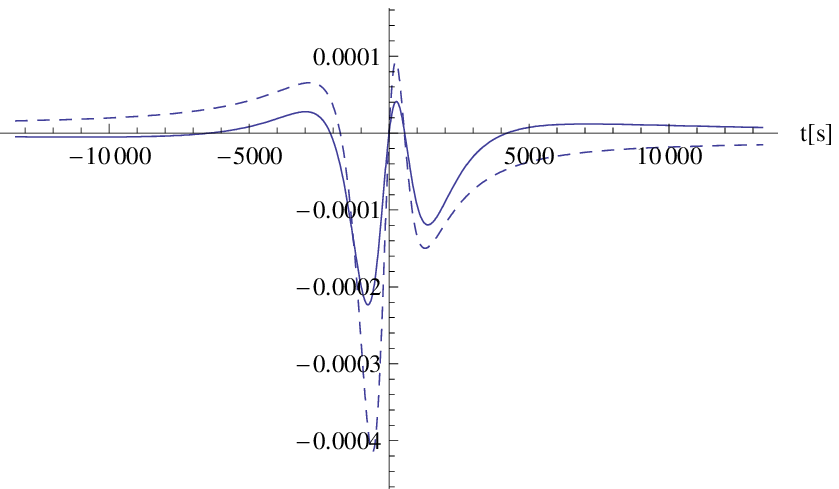}\qquad %
\includegraphics[scale=0.9]{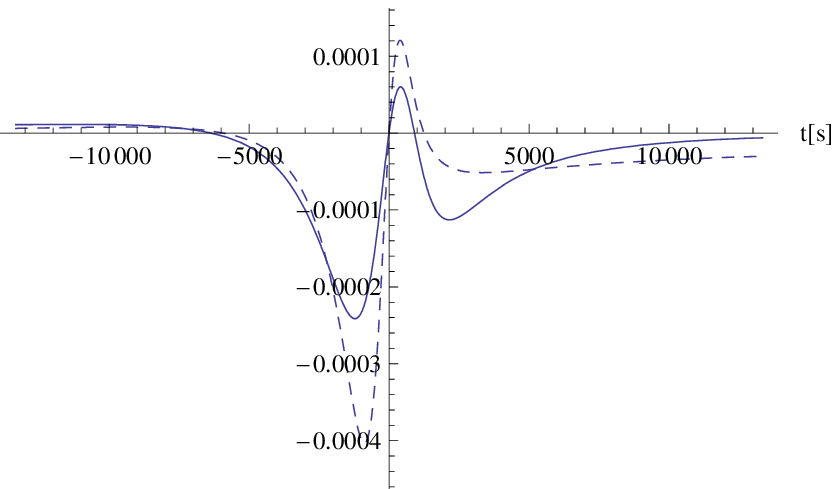}
\caption{Time dependence of the first corrections to the waveform
polarization states at 0.5PN order. Again, the corrections to the
$h_{+}$ polarization are shown on the left and for $h_{\times }$ on
the right.
The solid line shows the parabolic case and the dashed line shows $\protect\epsilon %
=2$ hyperbolic source. For further comparison amplitudes are multiplied by
the factor $10^{20}$.}
\label{fig:open05}
\end{figure}

\begin{figure}[th]
\noindent\hfil
\includegraphics[scale=0.9]{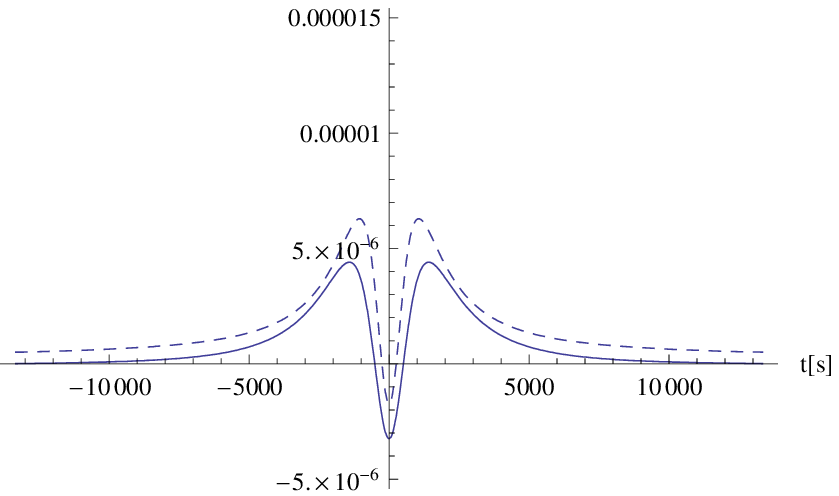}\qquad %
\includegraphics[scale=0.9]{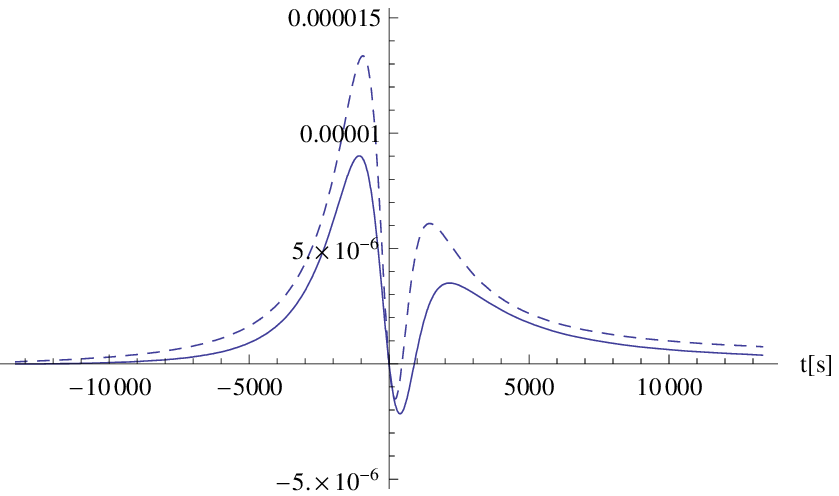}
\caption{Multipolar corrections to the polarization states at 1PN
order. Again, the corrections to $h_{+}$ are shown on the left and
to $h_{\times }$ on the right. Parabolic and hyperbolic orbit cases
are depicted by solid and dashed lines, respectively. Again, the
amplitudes are multiplied by the factor $10^{20}$.}
\label{fig:openPN}
\end{figure}

\begin{figure}[ht!]
\noindent\hfil
\includegraphics[scale=0.9]{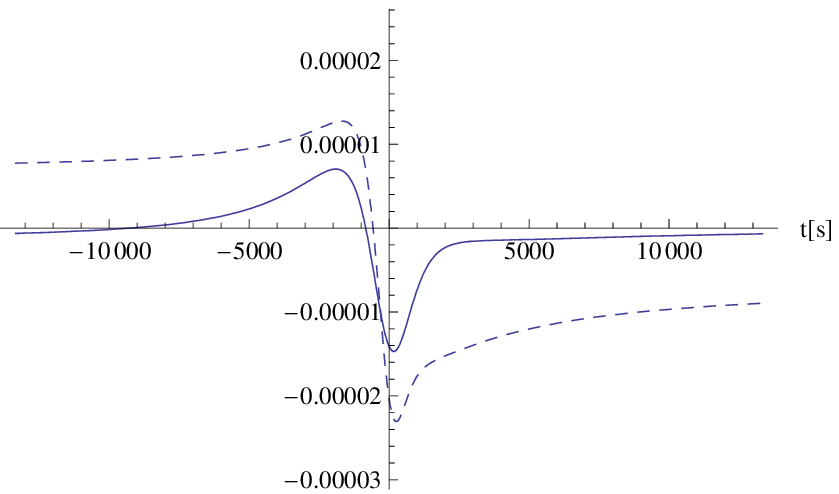}\qquad %
\includegraphics[scale=0.9]{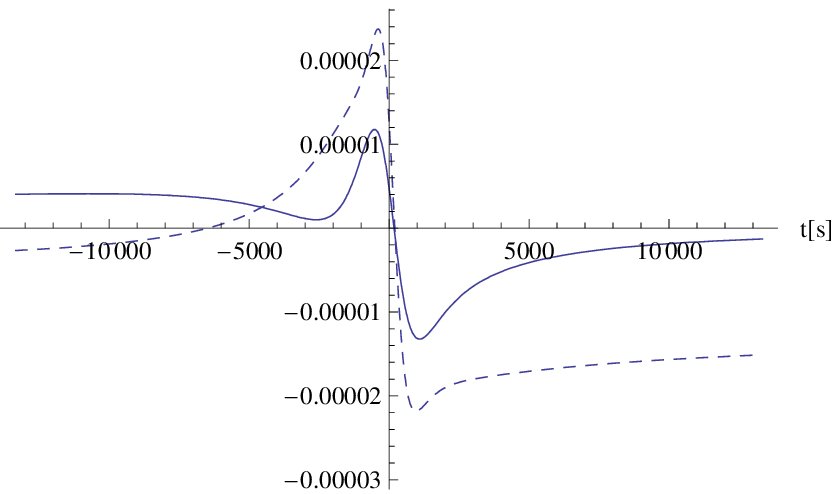}
\caption{1PN order corrections to the polarization states, which
arise from the perturbative description of the elements of the
motion. The notation is the same as on the former figures in this
subsection. To compare the different terms of these corrections, the
amplitudes are multiplied by the same factor as before, namely
$10^{20}$.} \label{fig:openNPN}
\end{figure}

\begin{figure}[th]
\noindent\hfil
\includegraphics[scale=0.9]{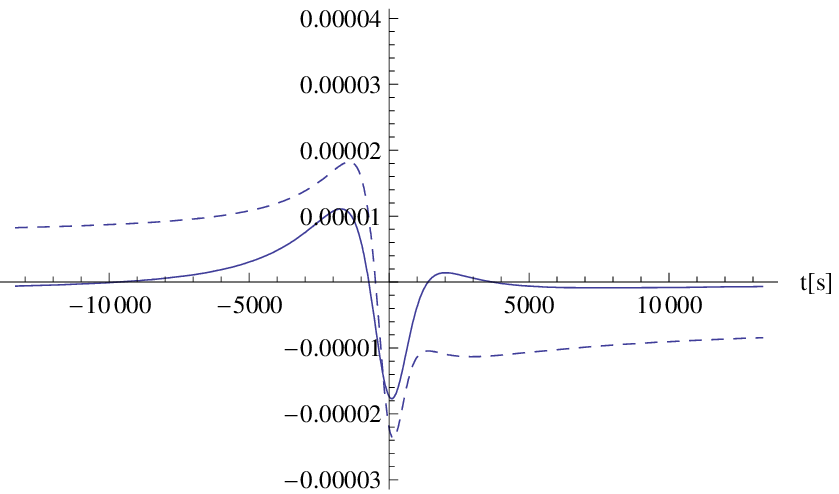}\qquad %
\includegraphics[scale=0.9]{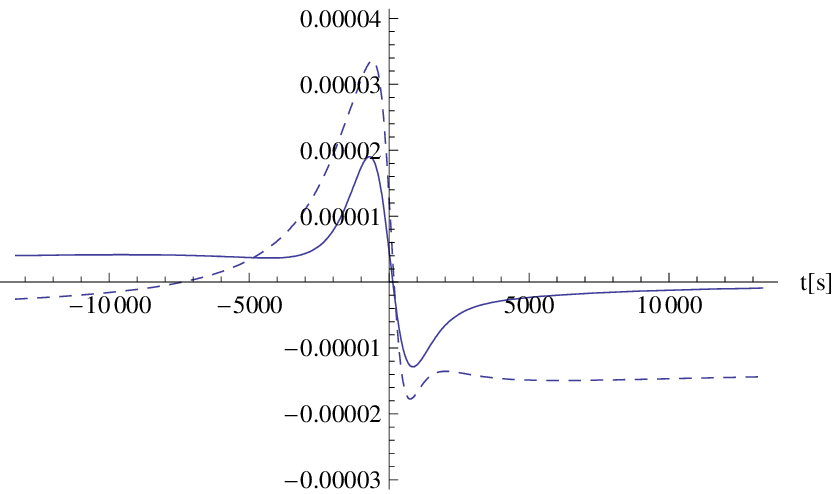}
\caption{In these figures we plot the full 1PN corrections to the
polarization states for the parabolic (soled line) and hyperbolic
(dashed line) sources. The corrections to the $h_{+}$ polarization
states are depicted on the left, while to $h_{\times }$ on the
right.} \label{fig:open1PN}
\end{figure}

In Figs. \ref{fig:openN}-\ref{fig:open1PN}, it can be seen that the
eccentricity of the orbit affects the frequency of the emitted
signal. With the same properties, the higher the eccentricity, the
more spiky the signal with higher amplitude at each PN order.

\section{Concluding remarks}

We have introduced a generalized true anomaly parametrization for
binary systems on either closed or on open orbit and applied it to
evaluate gravitational waveforms emitted by them. In addition to
elliptic orbits we consider the scattering of binary systems
following quasiparabolic and quasihyperbolic orbits. Our new
parametrization leads to a universal description valid for elliptic,
parabolic and hyperbolic cases. The only differences are in the
numerical integration determining the time dependence. With the use
of this parametrization the radial and angular equations of the
motion are solved, and with the inclusion of the results the
analytic form of the polarization states for the emitted
gravitational wave signals are evaluated for all types of sources
including circular orbits, too.

The time dependence of the generalized true anomaly parameter $\chi
$ is determined numerically for arbitrary parameter values. We have
evaluated polarization states of the emitted radiation for
realistic, and some well-known astrophysical sources. For elliptic
orbits we presented the corrections to the waveform order by order
for the Hulse-Taylor and J0737-3039 pulsars. For open orbits we have
computed the waveforms for two sources with physically reasonable
orbital elements to illustrate their main characteristics.

Our results illustrate the importance of the effects of eccentricity arising
from the presence of higher harmonics in the waveform even at Newtonian
order. The eccentricity increases the possibility of the detection of GWs
emitted by sources on such orbits even if the eccentricity is very small, as
it is in the case of the pulsar J0737-3039.

Our investigation of open orbit sources show that they provide
unique, burstlike wave signals, and this behavior becomes more
characteristic for higher order corrections. Meanwhile, these
sources and the emitted waveforms can be described analytically,
only the time dependence of the parameter has to be treated
numerically. It gives the opportunity to investigate and create data
banks for these sources to use matched filtering techniques.

Though the event rate estimations show that the detection of hyperbolic
orbits are less probable, the emitted signal is more "spiky" for hyperbolic
orbits than for parabolic ones with higher amplitude and frequency. These
differences become more significant at higher PN order. We have chosen a
quite large impact parameter (namely, almost equal to the periastron of the
Hulse-Taylor pulsar) to compare the differences between closed and open
orbit cases, and to investigate the effects of the eccentricity on the
widest possible range. With the application of the parameter values chosen
in \cite{KB}, the frequency and the amplitude of the emitted waveform
significantly increases.

The burstlike behavior of the waveform for open orbits, and the
possibility of an analytic description, make these systems an
important source of detectable gravitational waves for LISA. The
significantly smaller impact parameter choice makes these systems
potentially detectable for even the advanced ground-based detectors.
The higher order description gives rise to many additional
advantages, since it provides higher frequency terms for the matched
filtering evaluations without the assumption of extreme parameter
values for the astrophysical sources.

The results detailed above provide the possibility to generalize
this formalism to describe spin and quadrupole effects on the
detectable waveform. Furthermore, at higher orders it is possible to
describe the closure of the orbit controlled by radiation reaction.

\begin{acknowledgments}
This work was supported by OTKA Grant No. NI68228. We would like to
thank \'Arp\'ad Luk\'acs and Bal\'azs Mik\'oczi for their valuable
suggestions and help.
\end{acknowledgments}

\section*{APPENDIX A: NEGLECTING INTERFERENCE TERMS IN $\protect\chi(t)$}

When solving Eq.(\ref{dchidt}) for elliptic orbits, the Newtonian
and perturbative terms cannot be straightforwardly separated because
of the secular divergences. The easiest way to solve the
differential equation numerically is to solve the Newtonian and 1PN
order equations with either the integration of the differential
equations or with the inversion of the analytic solution $t(\chi )$.

When the different order contributions in the resulting numerical
data are to be separated it is nontrivial to neglect the
interference terms. In the following we give the description of the
problem, and give the guidelines for the solution.

The Newtonian order equation has the following structure
\begin{eqnarray}
\frac{d\chi_N}{dt}=f_N(\cos{\chi_N})\ ,
\end{eqnarray}
and similarly the 1PN order equation is
\begin{eqnarray}
\frac{d\chi}{dt}=f(\cos{\chi})\ ,
\end{eqnarray}
where
\begin{eqnarray}
f(\cos{\chi})=f_N(\cos{\chi})+f_P(\cos{\chi})\ ,  \notag
\end{eqnarray}
and we wish to find the solution of the form
\begin{eqnarray}
\chi_N=\Omega_Nt+g_N(t)\ ,\qquad \chi=(\Omega_N+\Omega_P)t+g(t)\ .
\end{eqnarray}
Using the above results one can evaluate both the $g_N(t)$ and
$g(t)$ functions ($\Omega_N$ and $\Omega_P$ are known from analytic
evaluations, see Sec. III.B.). To be able to determine the
contributions of the different order to the
\begin{eqnarray}
\chi=(\Omega_N+\Omega_P)t+\tilde{g}_N(t)+g_P(t)
\end{eqnarray}
solution, our main goal is to determine the function $\tilde{g}_N(t)$.

To achieve this we rewrite the above differential equations as
\begin{eqnarray}
\frac{dg_N(t)}{dt}=f_N(\cos{[\Omega_Nt+g_N(t)]})-\Omega_N  \label{dgNdt}
\end{eqnarray}
and
\begin{eqnarray}
\frac{d[\tilde{g}_N(t)+g_P(t)]}{dt}&=&f_N(\cos{[(\Omega_N+\Omega_P)t+\tilde{g%
}_N(t)]}- g_P(t)\sin{[(\Omega_N+\Omega_P)t+\tilde{g}_N(t)]})  \notag \\
&+&f_P(\cos{[(\Omega_N+\Omega_P)t+\tilde{g}_N(t)]}-(\Omega_N+\Omega_P)+
\mathit{o}(\epsilon^2)\ .  \notag
\end{eqnarray}
After introducing the
\begin{eqnarray}
t^{\prime }=\frac{\Omega_N-\Omega_P}{\Omega_N}t  \notag
\end{eqnarray}
variable we will analyze the equation
\begin{eqnarray}
\frac{d[\tilde{g}_N(t^{\prime })+g_P(t^{\prime })]}{dt}&=&f_N(\cos{%
[\Omega_Nt+\tilde{g}_N(t^{\prime })]}- g_P(t^{\prime })\sin{[\Omega_Nt+%
\tilde{g}_N(t^{\prime })]})\left(1-\frac{\Omega_P}{\Omega_N}\right)  \notag
\\
&+&f_P(\cos{[\Omega_Nt+\tilde{g}_N(t^{\prime })]}-\Omega_N+\mathit{o}%
(\epsilon^2)\ .  \notag
\end{eqnarray}
This equation will become
\begin{eqnarray}
\frac{d\tilde{g}_N(t^{\prime })}{dt}\Bigg|_N&=&f_N(\cos{[\Omega_Nt+\tilde{g}%
_N(t^{\prime })]}-\Omega_N
\end{eqnarray}
at the Newtonian order. Comparing it with Eq.(\ref{dgNdt}) it turns out that
\begin{eqnarray}
\tilde{g}(t^{\prime })=g_N(t)=g_N\left(\left[1+\frac{\Omega_P}{\Omega_N}%
\right]t^{\prime }\right)\ .  \notag
\end{eqnarray}
With the use of it the whole solution of the problem is
\begin{eqnarray}
\chi=(\Omega_N+\Omega_P)t+g_N\left(\left[1+\frac{\Omega_P}{\Omega_N}\right]%
t\right)+ \left(g(t)-g_N\left(\left[1+\frac{\Omega_P}{\Omega_N}\right]%
t\right)\right)\ ,
\end{eqnarray}
where $g_N(t)$ and $g(t)$ are to be determined numerically, $\Omega_N$ and $%
\Omega_P$ are evaluated analytically.

\section*{APPENDIX B: CONSTANT COEFFICIENTS FOR THE WAVEFORM EXPRESSIONS}

Our analytic calculations give rise to complicated expressions for
the polarization states of the gravitational wave signals in all
relevant cases. In the previous sections the main structure of the
waveforms was given and their main properties were analyzed. In this
Appendix we collect the explicit form of the constant coefficients
appearing in the waveform expressions.

We start by introducing the following short-hand notations:
\begin{eqnarray}
K&=&1-3\eta\ ,\qquad B=8-3\eta\ ,\qquad D=\frac{m\mu}{A_0}\ ,  \notag \\
A^N_1&=&\frac{A_0^2}{4L^2}\ ,\quad A^N_2=\frac{A_0m\mu}{8L^2}\ ,\quad A^N_3=%
\frac{m^2\mu^2}{2L^2}\ ,  \notag \\
A^N_4&=&\frac{A_0^3m\mu}{8L^4}\ ,\quad A^N_5=\frac{A_0^2m^2\mu^2}{16L^4}\
,\quad A^N_6=\frac{A_0m^3\mu^3}{4L^4}\ ,\quad A^N_7=\frac{E^2}{16\mu^2}\
,\quad A^N_8=\frac{m^2\mu E}{8L^2}\ ,\quad A^N_9=\frac{m^4\mu^4}{16L^4}\ ,
\notag \\
A^N_{10}&=&DA^N_7\ ,\quad A^N_{11}=DA^N_8\ ,\quad A^N_{12}=DA^N_9\ ,  \notag
\\
A^{0.5}_1&=&\frac{\delta mA_0^3}{256L^3m}\ ,\quad A^{0.5}_2=\frac{\delta
mA_0^2\mu}{128L^3}\ ,\quad A^{0.5}_3=\frac{\delta mA_0m\mu^2}{64L^3}\ ,\quad
A^{0.5}_4=\frac{\delta mm^2\mu^3}{32L^3}\ ,  \notag \\
A^1_1&=&\frac{A_0^4}{64L^4}\ ,\qquad A^1_2=A^N_4/384\ ,\qquad
A^1_3=A^N_5/48\ ,\qquad A^1_4=A^N_6/384\ ,\qquad A^1_5=A^N_9/3\ .
\label{Acoeffs}
\end{eqnarray}
Although the above coefficients are not independent, they are
introduced separately to keep our tables clear and well-defined.

As it can be seen from the results of Sec. VI. in the analytic
waveform expressions, Eqs.(\ref{helliptic}), the coefficients $C_F$
and $S_F$ of $\chi$ and $\chi'$ are time independent. These
coefficients depend on the different harmonics of the angle $\gamma$
and we introduce the following general decomposition valid at least
up to 1PN order and for both polarization states:
\begin{eqnarray}
C_F=\sum_{l=0}^{\infty}\left[a^{C_F}_l\sin{(l\gamma)}+b^{C_F}_l\cos{(l\gamma)}\right]\ ,\nonumber\\
S_F=\sum_{l=0}^{\infty}\left[a^{S_F}_l\sin{(l\gamma)}+b^{S_F}_l\cos{(l\gamma)}\right]\
.\label{CScoeff}
\end{eqnarray}
Here $F$ is a multi-index containing the index of the polarization
state, the PN order of the expression, and the index of the
harmonics of $\chi$ and/or $\chi'$. The coefficients $a^{C_F}_l$,
$b^{C_F}_l$, $a^{S_F}_l$ ,and $b^{S_F}_l$ in Eq.(\ref{CScoeff}) will
be given in tabular form for both polarization states and PN orders.
Any coefficients $a^{C_F}_l$, $b^{C_F}_l$, $a^{S_F}_l$ and
$b^{S_F}_l$ not listed in the tables are identically zero. With the
use of these coefficients listed in the tables the quantities $C_F$
and $S_F$ in Eq.(\ref{CScoeff}) are easily reconstructed.

\subsection{Circular orbit case}

Let us recall the results presented in Sec. VI.B. The polarization
states for the detectable gravitational wave signal can be written
as
\begin{eqnarray}
h^{N_{\times}^{+}}=\sum_{k=0}^{2}\big[C_k^{N_{\times}^{+}}\cos{%
(k\chi^{\prime})}+ S_k^{N_{\times}^{+}}\sin{(k\chi^{\prime})}\big]\ ,
\label{hNcirc}
\end{eqnarray}
\begin{eqnarray}
h^{0.5_{\times}^{+}}=\sum_{k=0}^{3}\big[C_k^{0.5_{\times}^{+}}\cos{%
(k\chi^{\prime})}+ S_k^{0.5_{\times}^{+}}\sin{(k\chi^{\prime})}\big]\ ,
\label{h05circ}
\end{eqnarray}
\begin{eqnarray}
h^{1_{\times}^{+}}&=&\sum_{k=0}^{4}\big[C_k^{1_{\times}^{+}}\cos{%
(k\chi^{\prime})}+ S_k^{1_{\times}^{+}}\sin{(k\chi^{\prime})}\big]\ ,
\label{h1circ}
\end{eqnarray}
where
\begin{eqnarray}
\chi^{\prime }=\Upsilon _{0}+\frac{Lt}{\mu r^2}-\frac{Lt}{\mu^2r^3}%
[(1-3\eta)Er+ (4-2\eta)m\mu]\ .  \notag
\end{eqnarray}

Instead of using the short-hand notations of Eq.(\ref{Acoeffs}), we
use the following expressions:
\begin{eqnarray}
a^N_1&=&\frac{L^2}{4r^2\mu^2}\ ,\qquad a^N_2=\frac{m}{4r}\ ,\qquad
a^N_3=(1-3\eta)\frac{EL^2}{2r^2\mu^3}\ ,\qquad a^N_4=(2-\eta)\frac{L^2m}{%
r^3\mu^2}\ ,  \notag \\
a^{0.5}_1&=&\frac{\delta mL^3}{32m\mu^3r^3}\ ,\qquad a^{0.5}_2=\frac{\delta
mL}{32\mu r^2}\ ,  \notag \\
a^1_1&=&\frac{L^4}{384\mu^4r^4}\ ,\qquad
a^1_2=\frac{L^2m}{384\mu^2r^3}\ ,\qquad a^1_3=\frac{m^2}{384r^2}\ .
\notag
\end{eqnarray}

As an example, the Newtonian expression for $C_{2}^{N+}$ arising
with the use of Table \ref{coeff:circNN} as
\begin{eqnarray}
C_{2}^{N+}=b^{C^{N+}_2}_2\cos{2\gamma}+b^{C^{N+}_2}_0=(a^N_1+a^N_2)\cos{2\gamma}+3(a^N_1+a^N_2)\ .
\end{eqnarray}
The relevant coefficients for circular orbits are listed in Tables
\ref{coeff:circNN} - \ref{coeff:circNPN}.

\begin{table}[!ht]
\begin{center}
\begin{tabular}{|c||c|c||c|}
\hline State & \multicolumn{2}{|c||}{$+$} & $\times$ \\ \hline
Harmonic & $\cos{2\gamma}$ & $1$ & $\cos{\gamma}$ \\
\hline\hline
$C_{2}^{N}$ & $a^N_1+a^N_2$ & $3(a^N_1+a^N_2)$ & $0$ \\ \hline
$C_{0}^{N}$ & $-a^N_1+a^N_2$ & $a^N_1-a^N_2$ & $0$ \\ \hline
$S_{2}^{N}$ & $0$ & $0$ & $4(a^N_1+a^N_2)$ \\ \hline
\end{tabular}%
\end{center}
\caption{This table contains the relevant $b^{C_k}_l$ and
$b^{S_k}_l$ coefficients for evaluating every $C^N_k$ and $S^N_k$ in
Eq.(\protect\ref{hNcirc}) restricted to the Newtonian order. Any
other coefficients are zero. The column under the sign "$+$"
contains the coefficients for $h^{N+}$, and under the sign
"$\times$" the coefficients for $h^{N\times}$ can be found.}
\label{coeff:circNN}
\end{table}

\begin{table}[!ht]
\begin{center}
\begin{tabular}{|c||c|c||c|}
\hline State & \multicolumn{2}{|c||}{$+$} & $\times$ \\ \hline
Harmonic & $\sin{3\gamma}$ & $\sin{\gamma}$ & $\sin{2\gamma}$ \\
\hline\hline
$C_{3}^{0.5}$ & $-4a^{0.5}_1$ & $-20a^{0.5}_1$ & $8a^{0.5}_1-28a^{0.5}_2$ \\
\hline
$C_{1}^{0.5}$ & $4a^{0.5}_1$ & $20a^{0.5}_1$ & $-8a^{0.5}_1-20a^{0.5}_2$ \\
\hline
$S_{3}^{0.5}$ & $-2a^{0.5}_1+7a^{0.5}_2$ & $5(-2a^{0.5}_1+7a^{0.5}_2)$ & $%
-16a^{0.5}_1$ \\ \hline
$S_{1}^{0.5}$ & $6a^{0.5}_1+7a^{0.5}_2$ & $-2a^{0.5}_1+19a^{0.5}_2$ & $%
16a^{0.5}_1$ \\ \hline
\end{tabular}%
\end{center}
\caption{This table details the relevant coefficients for evaluating
every $C^N_k$ and $S^N_k$ in Eq.(\protect\ref{h05circ}) for both
polarization states. The coefficients not included are zero. Again,
the column under the sign "$+$" contains the coefficients for
$h^{0.5+}$, and under the sign "$\times$" the coefficients for
$h^{0.5\times}$ can be found.} \label{coeff:circ05}
\end{table}

\begin{table}[!ht]
\begin{center}
\begin{tabular}{|c||c|c|c|}
\hline Harmonic & $\cos{4\gamma}$ & $\cos{2\gamma}$ & $1$ \\
\hline\hline
$C_{4}^{1}$ & $K(6a^1_1+51a^1_2+7a^1_3)$ & $4K(6a^1_1+51a^1_2+7a^1_3)$ & $%
5K(6a^1_1+51a^1_2+7a^1_3)$ \\ \hline
$C_{2}^{1}$ & $4K(-6a^1_1+3a^1_2+7a^1_3)$ & $16[3Ka^1_1+9(1+K)a^1_2-29a^1_3]$
& $4[42Ka^1_1+3(7+87\eta)a^1_2+(-355+21\eta)a^1_3]$ \\ \hline
$C_{0}^{1}$ & $3K(6a^1_1-13a^1_2+7a^1_3)$ & $12[-6Ka^1_1+(23-17%
\eta)a^1_2+(-41+7\eta)a^1_3]$ & $3[18Ka^1_1+(-79+29\eta)a^1_2+(157-7%
\eta)a^1_3]$ \\ \hline
\end{tabular}%
\end{center}
\caption{This table details the relevant coefficients for evaluating
every nonzero $C^1_k$ in Eq.(\protect\ref{h1circ}) for $h^{1+}$. It
is worth mentioning that in this case all $S^1_k$ are zero.}
\label{coeff:circPNplus}
\end{table}

\begin{table}[!ht]
\begin{center}
\begin{tabular}{|c||c|c|}
\hline Harmonic & $\cos{3\gamma}$ & $\cos{\gamma}$ \\ \hline\hline
$S_{4}^{1}$ & $4K(6a^1_1+51a^1_2+7a^1_3)$ & $4K(6a^1_1+51a^1_2+7a^1_3)$ \\
\hline
$S_{2}^{1}$ & $8K(-6a^1_1+15a^1_2+7a^1_3)$ & $8[30Ka^1_1+3(11+39%
\eta)a^1_2+(-239+21\eta)a^1_3]$ \\ \hline
\end{tabular}%
\end{center}
\caption{This table details the relevant coefficients for evaluating
every nonzero $S^1_k$ in Eq.(\protect\ref{h1circ}) for
$h^{1\times}$. It is worth mentioning that in this case all $C^1_k$
are zero.} \label{coeff:circPNcross}
\end{table}

\begin{table}[!ht]
\begin{center}
\begin{tabular}{|c||c|c||c|}
\hline State & \multicolumn{2}{|c||}{$+$} & $\times$ \\ \hline
Harmonic & $\cos{2\gamma}$ & $1$ & $\cos{\gamma}$ \\ \hline\hline
$C_{2}^{N}$ & $a^N_1+a^N_2-(a^N_3+a^N_4)$ &
$3(a^N_1+a^N_2)-3(a^N_3+a^N_4)$ & $0$ \\ \hline
$C_{0}^{N}$ & $-a^N_1+a^N_2+(a^N_3+a^N_4)$ & $a^N_1-a^N_2-(a^N_3+a^N_4)$ & $%
0 $ \\ \hline
$S_{2}^{N}$ & $0$ & $0$ & $4(a^N_1+a^N_2)-4(a^N_3+a^N_4)$ \\ \hline
\end{tabular}%
\end{center}
\caption{In this table the full coefficients are presented for Eq.(\protect\ref{hNcirc}),
including 1\,PN order terms from the corrections of the description of the motion. The table
is organized in the same way as Table \ref{coeff:circNN}.}
\label{coeff:circNPN}
\end{table}

\subsection{Elliptic orbit case}

Here we recall the results discussed in Sec. VI. A to collect all
the expressions for the elliptic orbit case:
\begin{eqnarray}
h^{N_{\times}^{+}} &=&\sum_{k=-2}^{2}\sum_{j=-3}^{3}\big[C_{k,j}^{N_{%
\times}^{+}}\cos {(k\chi ^{\prime }+j\chi ) }+S_{k,j}^{N_{\times}^{+}}\sin {%
(k\chi ^{\prime }+j\chi )}\big]\ ,  \label{hNell}
\end{eqnarray}
\begin{eqnarray}
h^{0.5_{\times}^{+}} &=&\sum_{k,j=-3}^{3}\big[C_{k,j}^{0.5_{\times}^{+}}\cos
{(k\chi ^{\prime }+j\chi )}+S_{k,j}^{0.5_{\times}^{+}}\sin {(k\chi ^{\prime
}+j\chi )}\big]\ ,  \label{h05ell}
\end{eqnarray}
\begin{eqnarray}
h^{1_{\times}^{+}} &=&\sum_{k,j=-4}^{4}\big[C_{k,j}^{1_{\times}^{+}}\cos {%
(k\chi ^{\prime }+j\chi ) }+S_{k,j}^{1_{\times}^{+}}\sin {(k\chi ^{\prime
}+j\chi )}\big]\ ,  \label{h1ell}
\end{eqnarray}
where
\begin{eqnarray}
\chi^{\prime }=\Upsilon _{0}+\chi-\frac{3m^{2}\mu ^{2}}{L^{2}}\chi\ .  \notag
\end{eqnarray}

Again, to collect the explicit expressions for the $C_{k,j}$ and $S_{k,j}$ constant
quantities we will use the short-hand notations and definitions of Eq.(\ref{Acoeffs})
and Eq.(\ref{CScoeff}).

For elliptic orbits our tables are organized in the same way as in
the circular orbit case. As an example, we detail the expression of
$C_{4,1}^{1+}$ with the use of Table \ref{coeff:ell1plus}:
\begin{eqnarray}
C_{4,1}^{1+}&=&b^{C^{1+}_{4,1}}_4\cos{[4\gamma]}+b^{C^{1+}_{4,1}}_2\cos{[2\gamma]}+b^{C^{1+}_{4,1}}_0\nonumber\\
&=&(K81A^1_2+K452A^1_4)\cos{[4\gamma]}+(324KA^1_2+1808KA^1_4)\cos{[2\gamma]}-
(405KA^1_2+2260KA^1_4)\ .
\end{eqnarray}
The relevant coefficients for elliptic orbits are listed in Tables
\ref{coeff:ellN} - \ref{coeff:ellPNcross}.

\begin{table}[!ht]
\begin{center}
\begin{tabular}{|c||c|c|}
\hline State & \multicolumn{2}{|c|}{$+$} \\ \hline Harmonic &
$\cos{2\gamma}$ & $1$ \\ \hline\hline $C_{2,1}^{N}$ & $A^N_2$ &
$3A^N_2$ \\ \hline $C_{2,0}^{N}$ & $A^N_3$ & $3A^N_3$ \\ \hline
$C_{2,-1}^{N}$ & $5A^N_2$ & $15A^N_2$ \\ \hline $C_{2,-2}^{N}$ &
$A^N_1$ & $3A^N_1$ \\ \hline $C_{0,1}^{N}$ & $-A^N_2$ & $A^N_2$ \\
\hline $C_{0,0}^{N}$ & $-A^N_1$ & $A^N_1$ \\ \hline
\end{tabular}
\qquad
\begin{tabular}{|c||c|}
\hline State & $\times$ \\ \hline Harmonic & $\cos{\gamma}$ \\
\hline\hline $S_{2,1}^{N}$ & $4A^N_2$ \\ \hline $S_{2,0}^{N}$ &
$4A^N_3$ \\ \hline $S_{-2,2}^{N}$ & $-4A^N_1$ \\ \hline
$S_{-2,1}^{N}$ & $-20A^N_2$ \\ \hline
\end{tabular}%
\end{center}
\caption{This table contains the relevant $b^{C_{k,j}}_l$ and $b^{S_{k,j}}_l$ coefficients
for evaluating every non-zero $C^N_{k,j}$ and $S^N_{k,j}$ in Eq.(\protect\ref{hNell}) restricted
to the Newtonian order. The left table contains the coefficients
for $h^{N+}$, and in the right one the coefficients for $h^{N\times}$ can be found.
It is worth to mention that for the "plus" state every $S^N_k$, and for the
"cross" state every $C^N_k$ are zero.}
\label{coeff:ellN}
\end{table}

\begin{table}[!ht]
\begin{center}
\begin{tabular}{|c||c|c||c|}
\hline State & \multicolumn{2}{|c||}{$+$} & $\times$ \\ \hline\hline
harmonic & $\sin{3\gamma}$ & $\sin{\gamma}$ & $\sin{2\gamma}$ \\
\hline\hline $C_{3,3}^{0.5}$ & $-4A^{0.5}_1$ & $-20A^{0.5}_1$ &
$-4A^{0.5}_1$ \\ \hline $C_{3,2}^{0.5}$ & $-12A^{0.5}_2$ &
$-60A^{0.5}_2$ & $-4A^{0.5}_2$ \\ \hline $C_{3,1}^{0.5}$ &
$-4(3A^{0.5}_1+A^{0.5}_3)$ & $-4(15A^{0.5}_1+A^{0.5}_3)$ &
$4(11A^{0.5}_1-2A^{0.5}_3)$ \\ \hline $C_{3,0}^{0.5}$ &
$-4(6A^{0.5}_2+A^{0.5}_4)$ & $-20(6A^{0.5}_2+A^{0.5}_4)$ &
$20(2A^{0.5}_2-A^{0.5}_4)$ \\ \hline $C_{3,-1}^{0.5}$ &
$-12(A^{0.5}_1+A^{0.5}_3)$ & $-60(A^{0.5}_1+A^{0.5}_3)$ &
$4(13A^{0.5}_1-14A^{0.5}_3)$ \\ \hline $C_{3,-2}^{0.5}$ &
$-12A^{0.5}_2$ & $-60A^{0.5}_2$ & $-60A^{0.5}_2$ \\ \hline
$C_{3,-3}^{0.5}$ & $-4A^{0.5}_1$ & $-20A^{0.5}_1$ & $-28A^{0.5}_1$
\\ \hline $C_{1,3}^{0.5}$ & $4A^{0.5}_1$ & $4A^{0.5}_1$ &
$-4A^{0.5}_1$ \\ \hline $C_{1,2}^{0.5}$ & $12A^{0.5}_2$ &
$60A^{0.5}_2$ & $-20A^{0.5}_2$ \\ \hline
$C_{1,1}^{0.5}$ & $12(A^{0.5}_1+A^{0.5}_3)$ & $60(A^{0.5}_1+A^{0.5}_3)$ & $%
-20(A^{0.5}_1+2A^{0.5}_3)$ \\ \hline
$C_{1,0}^{0.5}$ & $4(6A^{0.5}_2+A^{0.5}_4)$ & $20(6A^{0.5}_2+A^{0.5}_4)$ & $%
-4(18A^{0.5}_2+7A^{0.5}_4)$ \\ \hline
$C_{1,-1}^{0.5}$ & $12(A^{0.5}_1+A^{0.5}_3)$ & $60(A^{0.5}_1+A^{0.5}_3)$ & $%
-4(3A^{0.5}_1+22A^{0.5}_3)$ \\ \hline
$C_{1,-2}^{0.5}$ & $12A^{0.5}_2$ & $60A^{0.5}_2$ & $-84A^{0.5}_2$ \\ \hline
$C_{1,-3}^{0.5}$ & $4A^{0.5}_1$ & $20A^{0.5}_1$ & $-28A^{0.5}_1$ \\ \hline
$S_{3,3}^{0.5}$ & $A^{0.5}_1$ & $5A^{0.5}_1$ & $-16A^{0.5}_1$ \\ \hline
$S_{3,2}^{0.5}$ & $-A^{0.5}_2$ & $-5A^{0.5}_2$ & $-48A^{0.5}_2$ \\ \hline
$S_{3,1}^{0.5}$ & $-11A^{0.5}_1+2A^{0.5}_3$ & $-5(11A^{0.5}_1-2A^{0.5}_3)$ &
$-48(A^{0.5}_1+A^{0.5}_3)$ \\ \hline
$S_{3,0}^{0.5}$ & $-5(2A^{0.5}_2-A^{0.5}_4)$ & $-25(2A^{0.5}_2-A^{0.5}_4)$ &
$-16(6A^{0.5}_2+A^{0.5}_4)$ \\ \hline
$S_{1,3}^{0.5}$ & $11A^{0.5}_1$ & $-25A^{0.5}_1$ & $16A^{0.5}_1$ \\ \hline
$S_{1,2}^{0.5}$ & $23A^{0.5}_2$ & $-29A^{0.5}_2$ & $48A^{0.5}_2$ \\ \hline
$S_{1,1}^{0.5}$ & $-23A^{0.5}_1-22A^{0.5}_3$ & $-29A^{0.5}_1+14A^{0.5}_3$ & $%
48(A^{0.5}_1+A^{0.5}_3)$ \\ \hline
$S_{1,0}^{0.5}$ & $38A^{0.5}_2+13A^{0.5}_4$ & $-30A^{0.5}_2-17A^{0.5}_4$ & $%
16(6A^{0.5}_2+A^{0.5}_4)$ \\ \hline
$S_{-1,3}^{0.5}$ & $-13A^{0.5}_1$ & $-17A^{0.5}_1$ & $-16A^{0.5}_1$ \\ \hline
$S_{-1,2}^{0.5}$ & $-39A^{0.5}_2$ & $-51A^{0.5}_2$ & $-48A^{0.5}_2$ \\ \hline
$S_{-1,1}^{0.5}$ & $-A^{0.5}_1-93/2A^{0.5}_3$ & $-21A^{0.5}_1-50A^{0.5}_3$ &
$-16(A^{0.5}_1+A^{0.5}_3)$ \\ \hline
$S_{-3,3}^{0.5}$ & $-7A^{0.5}_1$ & $-35A^{0.5}_1$ & $16A^{0.5}_1$ \\ \hline
$S_{-3,2}^{0.5}$ & $-15A^{0.5}_2$ & $-75A^{0.5}_2$ & $48A^{0.5}_2$ \\ \hline
$S_{-3,1}^{0.5}$ & $13A^{0.5}_1-14A^{0.5}_3$ & $5(13A^{0.5}_1-14A^{0.5}_3)$
& $48(A^{0.5}_1+A^{0.5}_3)$ \\ \hline
\end{tabular}%
\end{center}
\caption{This table contains the relevant $a^{C_{k,j}}_l$ and $a^{S_{k,j}}_l$ coefficients
for evaluating the explicit form of Eq.(\protect\ref{h05ell}). Any other coefficients
are zero in this equation. The columns under
the sign "$+$" contains the coefficients for $h^{0.5+}$
and under the sign "$\times$" the coefficients for $h^{0.5\times}$ can be found.}
\label{coeff:ell05}
\end{table}

\begin{table}[!ht]
\begin{center}
\begin{tabular}{|c||c|c|c|}
\hline Harmonic & $\cos{[4\gamma]}$ & $\cos{[2\gamma]}$ & $1$ \\
\hline\hline $C_{4,3}^{1+}$ & $15KA^1_2$ & $60KA^1_2$ & $-75KA^1_2$
\\ \hline $C_{4,2}^{1+}$ & $35KA^1_3$ & $140KA^1_3$ & $-175KA^1_3$
\\ \hline
$C_{4,1}^{1+}$ & $K81A^1_2+K452A^1_4$ & $324KA^1_2+1808KA^1_4$ & $%
-405KA^1_2-2260KA^1_4$ \\ \hline
$C_{4,0}^{1+}$ & $142KA^1_3+16KA^1_5$ & $568KA^1_3+32KA^1_5$ & $%
-710KA^1_3-40KA^1_5$ \\ \hline
$C_{4,-1}^{1+}$ & $189KA^1_2+1076KA^1_4$ & $756KA^1_2+4304KA^1_4$ & $%
-945KA^1_2-5380KA^1_4$ \\ \hline
$C_{4,-2}^{1+}$ & $215KA^1_3$ & $860KA^1_3$ & $-1075KA^1_3$ \\ \hline
$C_{4,-3}^{1+}$ & $315KA^1_2$ & $1260KA^1_2$ & $-1575KA^1_2$ \\ \hline
$C_{4,-4}^{1+}$ & $KA^1_1$ & $4KA^1_1$ & $-5KA^1_1$ \\ \hline\hline
$C_{2,3}^{1+}$ & $-24KA^1_2$ & $96(2-9\eta)A^1_2$ & $24(1-39\eta)A^1_2$ \\
\hline
$C_{2,2}^{1+}$ & $-40KA^1_3$ & $8(4-123\eta)A^1_3$ & $-8(103+243\eta)A^1_3$
\\ \hline
$C_{2,1}^{1+}$ & $-144KA^1_2-256KA^1_4$ & $96(5-14\eta)A^1_2-64(13+144%
\eta)A^1_4$ & $48(39\eta-11)A^1_2+64(42\eta-167)A^1_4$ \\ \hline
$C_{2,0}^{1+}$ & $-152KA^1_3+2KA^1_5$ & $16(34-123\eta)A^1_3-8(2+9\eta)A^1_5$
& $8(69\eta-65)A^1_3+2(39\eta-73)A^1_5$ \\ \hline
$C_{2,-1}^{1+}$ & $-504KA^1_2+224KA^1_4$ & $96(45\eta-22)A^1_2-64(29-120%
\eta)A^1_4$ & $8(543-873\eta)A^1_2+32(237\eta-145)A^1_4$ \\ \hline
$C_{2,-2}^{1+}$ & $-4KA^1_1+32KA^1_3$ & $8KA^1_1+8(100-171\eta)A^1_3$ & $%
28KA^1_1+336(128+3\eta)A^1_3$ \\ \hline
$C_{2,-3}^{1+}$ & $0$ & $96(11-16\eta)A^1_2$ & $96(23-18\eta)A^1_2$ \\ \hline
$C_{0,3}^{1+}$ & $30KA^1_2$ & $24(29-15\eta)A^1_2$ & $6(75\eta-121)A^1_2$ \\
\hline
$C_{0,2}^{1+}$ & $-30KA^1_3$ & $120(1+3\eta)A^1_3$ & $-30(5+9\eta)A^1_3$ \\
\hline
$C_{0,1}^{1+}$ & $234KA^1_2-24KA^1_4$ & $24(235\eta-33)A^1_2-96(37-35%
\eta)A^1_4$ & $6(93-823\eta)A^1_2+24(149-143\eta)A^1_4$ \\ \hline
$C_{0,0}^{1+}$ & $3KA^1_1-6KA^1_3$ & $-12KA^1_1+24(13-49\eta)A^1_3+12(3-%
\eta)A^1_5$ & $9KA^1_1+6(53-199\eta)A^1_3+12(3-\eta)A^1_5$ \\ \hline
\end{tabular}%
\end{center}
\caption{This table details the relevant coefficients for evaluating
every nonzero $C_{k,j}$ in Eq.(\protect\ref{h1ell}) for $h^{1+}$. It
is worth mentioning that for $h^{1+}$ every $S_{k,j}$ are zero.}
\label{coeff:ell1plus}
\end{table}

\begin{table}[!ht]
\begin{center}
\begin{tabular}{|c||c|c|}
\hline Harmonic & $\cos{[3\gamma]}$ & $\cos{[\gamma]}$ \\
\hline\hline $S_{4,3}^{1\times}$ & $60KA^1_2$ & $-60KA^1_2$ \\
\hline $S_{4,2}^{1\times}$ & $140KA^1_3$ & $-140KA^1_3$ \\ \hline
$S_{4,1}^{1\times}$ & $324KA^1_2+1808KA^1_4$ & $-324KA^1_2+1808KA^1_4$ \\
\hline
$S_{4,0}^{1\times}$ & $568KA^1_3+32KA^1_5$ & $-568KA^1_3+32KA^1_5$ \\
\hline\hline
$S_{2,3}^{1\times}$ & $24KA^1_2$ & $24(7-69\eta)A^1_2$ \\ \hline
$S_{2,2}^{1\times}$ & $40KA^1_3$ & $-8(109+117\eta)A^1_3$ \\ \hline
$S_{2,1}^{1\times}$ & $-24KA^1_2+544KA^1_4$ & $24(37\eta-7)A^1_2-32(585+69%
\eta)A^1_4$ \\ \hline
$S_{2,0}^{1\times}$ & $16KA^1_3-16KA^1_5$ & $-112(1+9\eta)A^1_3-16(11-3%
\eta)A^1_5$ \\ \hline\hline
$S_{0,3}^{1\times}$ & $-48KA^1_2$ & $48KA^1_2$ \\ \hline
$S_{0,2}^{1\times}$ & $-48KA^1_3$ & $48KA^1_3$ \\ \hline
$S_{0,1}^{1\times}$ & $-48KA^1_2+192KA^1_4$ & $48KA^1_2+192KA^1_4$ \\
\hline\hline
$S_{-2,3}^{1\times}$ & $-120KA^1_2$ & $24(121\eta-131)A^1_2$ \\ \hline
$S_{-2,2}^{1\times}$ & $8KA^1_1-232KA^1_3$ & $-40KA^1_1-8(203-93\eta)A^1_3$
\\ \hline
$S_{-2,1}^{1\times}$ & $672KA^1_2-1696KA^1_4$ & $24(495%
\eta-277)A^1_2+32(133-135\eta)A^1_4$ \\ \hline\hline
$S_{-4,4}^{1\times}$ & $-4KA^1_1$ & $4KA^1_1$ \\ \hline
$S_{-4,3}^{1\times}$ & $-1260KA^1_2$ & $1260KA^1_2$ \\ \hline
$S_{-4,2}^{1\times}$ & $-860KA^1_3$ & $860KA^1_3$ \\ \hline
$S_{-4,1}^{1\times}$ & $-756KA^1_2-4304KA^1_4$ & $756KA^1_2+4304KA^1_4$ \\
\hline
\end{tabular}%
\end{center}
\caption{This table details the relevant coefficients for evaluating
every nonzero $S_{k,j}$ in Eq.(\protect\ref{h1ell}) for
$h^{1\times}$. It is worth mentioning that for $h^{1\times}$ every
$C_{k,j}$ are zero.} \label{coeff:ell1cross}
\end{table}

Tables \ref{coeff:ellPNplus} and \ref{coeff:ellPNcross} contain the
1PN corrections to the Newtonian coefficients given in Table
\ref{coeff:ellN}, from which the full coefficients in
Eq.(\ref{hNell}) can be reconstructed.

\begin{table}[!ht]
\begin{center}
\begin{tabular}{|c||c|c|}
\hline Harmonic & $\cos{[2\gamma]}$ & $1$ \\ \hline\hline
$C_{2,2}^{N+}$ & $-BA^N_5$ & $-3BA^N_5$ \\ \hline
$C_{2,1}^{N+}$ & $(13\eta-31)A^N_{10}-(10-9\eta)A^N_{11}-2(14-17%
\eta)A^N_{12} $ & $3(13\eta-31)A^N_{10}-3(10-9\eta)A^N_{11}-6(14-17%
\eta)A^N_{12}$ \\ \hline
$C_{2,0}^{N+}$ & $2(5\eta-3)A^N_8+8(5\eta-2)A^N_9$ & $6(5\eta-3)A^N_8+24(5%
\eta-2)A^N_9$ \\ \hline
$C_{2,-1}^{N+}$ & $(65\eta-51)A^N_{10}-(32-43\eta)A^N_{11}-2(18-23%
\eta)A^N_{12}$ & $3(65\eta-51)A^N_{10}-3(32-43\eta)A^N_{11}-6(18-23%
\eta)A^N_{12}$ \\ \hline
$C_{2,-2}^{N+}$ & $-12KA^N_7+(33\eta-32)A^N_8+(53\eta-48)A^N_9$ & $%
-36KA^N_7+3(33\eta-32)A^N_8+3(53\eta-48)A^N_9$ \\ \hline
$C_{2,-3}^{N+}$ & $(3\eta-4)A^N_4$ & $3(3\eta-4)A^N_4$ \\ \hline
$C_{0,3}^{N+}$ & $2(2-\eta)A^N_4$ & $6(\eta-2)A^N_4$ \\ \hline
$C_{0,2}^{N+}$ & $(5-4\eta)A^N_8+4(12-11\eta)A^N_9$ & $4(4%
\eta-5)A^N_8-4(12-11\eta)A^N_9$ \\ \hline
$C_{0,1}^{N+}$ & $2(43-49\eta)A^N_{10}+2(41-34\eta)A^N_{11}+4(28-25%
\eta)A^N_{12}$ & $2(49\eta-43)A^N_{10}-2(41-34\eta)A^N_{11}-4(28-25%
\eta)A^N_{12}$ \\ \hline
$C_{0,0}^{N+}$ & $12KA^N_7+2(21-19\eta)A^N_8+12(6-5\eta)A^N_9$ & $%
-12KA^N_7-2(21-19\eta)A^N_8-12(6-5\eta)A^N_9$ \\ \hline\hline
$S_{2,0}^{N+}$ & $-4BA^N_5$ & $-12BA^N_5$ \\ \hline
$S_{0,1}^{N+}$ & $2BA^N_4$ & $-2BA^N_4$ \\ \hline
$S_{-2,1}^{N+}$ & $2BA^N_4$ & $6BA^N_4$ \\ \hline
\end{tabular}%
\end{center}
\caption{1PN corrections to the coefficients for $h^{N+}$ in
Eq.(\protect\ref{hNell}). The table contains all coefficients to
evaluate each nonzero $C_{k,j}$ and $S_{k,j}$. Any other
coefficients for these expressions are zero.}
\label{coeff:ellPNplus}
\end{table}

\begin{table}[!ht]
\begin{center}
\begin{tabular}{|c||c|}
\hline Harmonic & $\cos{[\gamma]}$ \\ \hline\hline
$C_{2,0}^{N\times}$ & $16BA^N_5$ \\ \hline $C_{2,-1}^{N\times}$ &
$8BA^N_4$ \\ \hline\hline $S_{2,2}^{N\times}$ & $-4BA^N_5$ \\ \hline
$S_{2,1}^{N\times}$ & $4(13\eta-31)A^N_{10}-4(10-9\eta)A^N_{11}-8(14-17%
\eta)A^N_{12}$ \\ \hline
$S_{2,0}^{N\times}$ & $8(5\eta-3)A^N_8-32(2-5\eta)A^N_9)$ \\ \hline
$S_{-2,3}^{N\times}$ & $4(4-3\eta)A^N_4$ \\ \hline
$S_{-2,2}^{N\times}$ & $48KA^N_7+4(32-33\eta)A^N_8+4(48-53\eta)A^N_9$ \\
\hline
$S_{-2,1}^{N\times}$ & $4(51-65\eta)A^N_{10}+4(32-43\eta)A^N_{11}+8(18-23%
\eta)A^N_{12}$ \\ \hline
\end{tabular}%
\end{center}
\caption{1PN corrections to the coefficients for $h^{N\times}$ in
Eq.(\protect\ref{hNell}). The table contains all coefficients to
evaluate each nonzero $S_{k,j}$ and $C_{k,j}$. Any other
coefficients for these expressions are zero.}
\label{coeff:ellPNcross}
\end{table}

\subsection{Open orbit case}

For open orbits the structure of the polarization states of the
gravitational waves can be written as
\begin{eqnarray}
h^{N_{\times}^{+}}=\chi\sum_{k=0}^{3}\big[C_{k\chi}^{N_{\times}^{+}}\cos{%
(k\chi)}+ S_{k\chi}^{N_{\times}^{+}}\sin{(k\chi)}\big]+ \sum_{k=0}^{4}\big[%
C_k^{N_{\times}^{+}}\cos{(k\chi)}+ S_k^{N_{\times}^{+}}\sin{(k\chi)}\big]\ ,
\label{hNopen}
\end{eqnarray}
\begin{eqnarray}
h^{0.5_{\times}^{+}}=\sum_{k=0}^{6}\big[C_k^{0.5_{\times}^{+}}\cos{(k\chi)}+
S_k^{0.5_{\times}^{+}}\sin{(k\chi)}\big]\ ,  \label{h05open}
\end{eqnarray}
\begin{eqnarray}
h^{1_{\times}^{+}}&=&\sum_{k=0}^{7}\big[C_k^{1_{\times}^{+}}\cos{(k\chi)}+
S_k^{1_{\times}^{+}}\sin{(k\chi)}\big]\ .  \label{h1open}
\end{eqnarray}

In this case there are no secular terms in the approximation scheme,
and hence a more straightforward series expansion can be given. It
makes the structure of the waveform expressions simpler, but on the
other hand the expressions for each coefficient become significantly
more complex. It is because in addition to the harmonics of the
angle $\gamma$ there appear the harmonics of $\Upsilon_0$, too.

To keep our tables simple we introduce the following short-hand
notations:
\begin{eqnarray}
C_{k}^{N+}&=&C_{kC2}^{N+}\cos{[2\gamma]}+C_{k0}^{N+}\ , \qquad\qquad\qquad\qquad
S_{k}^{N+}=S_{kC2}^{N+}\cos{[2\gamma]}+S_{k0}^{N+} \nonumber\\
C_{k}^{N\times}&=&C_{kC1}^{N\times}\cos{[\gamma]}\ , \qquad\qquad\qquad\qquad\qquad\qquad
S_{k}^{N\times}=S_{kC1}^{N\times}\cos{[\gamma]} \nonumber\\
C_{k}^{0.5+}&=&C_{kS3}^{0.5+}\sin{[3\gamma]}+C_{kS1}^{0.5+}\sin{[\gamma]}\ , \quad\qquad\qquad
S_{k}^{0.5+}=S_{kS3}^{0.5+}\sin{[3\gamma]}+S_{kS1}^{0.5+}\sin{[\gamma]} \nonumber\\
C_{k}^{0.5\times}&=&C_{kS2}^{0.5\times}\sin{[2\gamma]}\ , \quad\qquad\qquad\qquad\qquad\qquad
S_{k}^{0.5\times}=S_{kS2}^{0.5\times}\sin{[2\gamma]} \nonumber\\
C_k^{1+}&=&C_{kC4}^{1+}\cos{[4\gamma]}+C_{kC2}^{1+}\cos{[2\gamma]}
+C_{k0}^{1+}\ , \qquad
S_k^{1+}=S_{kC4}^{1+}\cos{[4\gamma]}+S_{kC2}^{1+}\cos{[2\gamma]}
+S_{k0}^{1+} \nonumber\\
C_k^{1\times}&=&C_{kC3}^{1\times}\cos{[3\gamma]}+
C_{kC1}^{1\times}\cos{[\gamma]}\ , \quad\qquad\qquad
S_k^{1\times}=S_{kC3}^{1\times}\cos{[3\gamma]}+
S_{kC1}^{1\times}\cos{[\gamma]}\ .
\end{eqnarray}

For open orbits we separate the higher harmonics in $\Upsilon_0$
instead of $\gamma$, and hence we write
\begin{eqnarray}
C_F=\sum_{l=0}^{\infty}\left[a^{C_F}_l\sin{(l\Upsilon_0)}+b^{C_F}_l\cos{(l\Upsilon_0)}\right]\ ,\nonumber\\
S_F=\sum_{l=0}^{\infty}\left[a^{S_F}_l\sin{(l\Upsilon_0)}+b^{S_F}_l\cos{(l\Upsilon_0)}\right]\ ,\label{CScoeffopen}
\end{eqnarray}
where $F$ is also a multi-index as for the elliptic case containing
also the index of the harmonics of $\gamma$. Tables
\ref{coeff:openPNN}-\ref{coeff:openPNchi} contain all the relevant
$a^{C_F}_l$, $b^{C_F}_l$, $a^{S_F}_l$ and $b^{S_F}_l$ coefficients.
Any other $a_l$ or $b_l$ coefficients are zero.

\begin{table}[!ht]
\begin{center}
\begin{tabular}{|c||c|c|c|}
\hline
Harmonic & $\cos{2\Upsilon_0}$ & $\sin{2\Upsilon_0}$ & 1 \\
\hline\hline
$C_{3C2}^{N}$ & $A^N_2$ & $0$ & $0$ \\ \hline
$C_{2C2}^{N}$ & $A^N_3$ & $0$ & $0$ \\ \hline
$C_{1C2}^{N}$ & $5A^N_2$ & $0$ & $-2A^N_2$ \\ \hline
$C_{0C2}^{N}$ & $A^N_1$ & $0$ & $-A^N_1$ \\
\hline\hline
$C_{30}^{N}$ & $3A^N_2$ & $0$ & $0$ \\ \hline
$C_{20}^{N}$ & $3A^N_3$ & $0$ & $0$ \\ \hline
$C_{10}^{N}$ & $15A^N_2$ & $0$ & $2A^N_2$ \\ \hline
$C_{00}^{N}$ & $3A^N_1$ & $0$ & $A^N_1$ \\
\hline\hline
$S_{3C2}^{N}$ & $0$ & $-A^N_2$ & $0$ \\ \hline
$S_{2C2}^{N}$ & $0$ & $-A^N_3$ & $0$ \\ \hline
$S_{1C2}^{N}$ & $0$ & $-5A^N_2$ & $0$ \\
\hline\hline
$S_{30}^{N}$ & $0$ & $-3A^N_2$ & $0$ \\ \hline
$S_{20}^{N}$ & $0$ & $-3A^N_3$ & $0$ \\ \hline
$S_{10}^{N}$ & $0$ & $-15A^N_2$ & $0$ \\
\hline
\end{tabular}%
\qquad
\begin{tabular}{|c||c|c|}
\hline
Harmonic & $\cos{2\Upsilon_0}$ & $\sin{2\Upsilon_0}$ \\
\hline\hline
$C_{3C1}^{N}$ & $0$ & $4A^N_2$ \\ \hline
$C_{2C1}^{N}$ & $0$ & $4A^N_3$ \\ \hline
$C_{1C1}^{N}$ & $0$ & $20A^N_2$ \\ \hline
$C_{0C1}^{N}$ & $0$ & $4A^N_1$ \\
\hline\hline
$S_{3C1}^{N}$ & $4A^N_2$ & $0$ \\ \hline
$S_{2C1}^{N}$ & $4A^N_3$ & $0$ \\ \hline
$S_{1C1}^{N}$ & $20A^N_2$ & $0$ \\
\hline
\end{tabular}%
\end{center}
\caption{This table contains the Newtonian value of all relevant
$a^{C_F}_l$, $a^{S_F}_l$, $b^{C_F}_l$, and $b^{S_F}_l$ coefficients
needed for evaluating every nonzero $C^N_F$ and $S^N_F$ in
Eq.(\protect\ref{hNopen}). The left table contains the coefficients
for $h^{N+}$, and in the right one the coefficients for
$h^{N\times}$ can be found.} \label{coeff:openN}
\label{coeff:openPNN}
\end{table}

\begin{table}[!ht]
\begin{center}
\begin{tabular}{|c||c|c|c|c|}
\hline Harmonic & $\cos{[3\Upsilon_0]}$ & $\cos{[\Upsilon_0]}$ &
$\sin{[3\Upsilon_0]}$ & $\sin{[\Upsilon_0]}$ \\ \hline\hline
$C_{6S3}^{0.5+}$ & $-4A^{0.5}_1$ & $0$ & $A^{0.5}_1$ & $0$ \\ \hline
$C_{5S3}^{0.5+}$ & $-12A^{0.5}_2$ & $0$ & $-A^{0.5}_2$ & $0$ \\
\hline
$C_{4S3}^{0.5+}$ & $-12(A^{0.5}_1+A^{0.5}_3)$ & $4A^{0.5}_1$ & $%
-11A^{0.5}_1+2A^{0.5}_3$ & $11A^{0.5}_1$ \\ \hline
$C_{3S3}^{0.5+}$ & $-4(6A^{0.5}_2+A^{0.5}_4)$ & $12A^{0.5}_2$ & $%
-5(2A^{0.5}_2-A^{0.5}_4)$ & $23A^{0.5}_2$ \\ \hline
$C_{2S3}^{0.5+}$ & $-12(A^{0.5}_1+A^{0.5}_3)$ & $4(4A^{0.5}_1+3A^{0.5}_3)$ & $%
-13A^{0.5}_1+14A^{0.5}_3$ & $2(28A^{0.5}_1+11A^{0.5}_3)$ \\ \hline
$C_{1S3}^{0.5+}$ & $-12A^{0.5}_2$ & $4(9A^{0.5}_2+A^{0.5}_4)$ & $15A^{0.5}_2$
& $77A^{0.5}_2+13A^{0.5}_4$ \\ \hline
$C_{0S3}^{0.5+}$ & $-4A^{0.5}_1$ & $12(A^{0.5}_1+A^{0.5}_3)$ & $7A^{0.5}_1$ & $%
A^{0.5}_1+42A^{0.5}_3$ \\
\hline\hline
$C_{6S1}^{0.5+}$ & $-20A^{0.5}_1$ & $0$ & $5A^{0.5}_1$ & $0$ \\ \hline
$C_{5S1}^{0.5+}$ & $-60A^{0.5}_2$ & $0$ & $-5A^{0.5}_2$ & $0$ \\ \hline
$C_{4S1}^{0.5+}$ & $-60(A^{0.5}_1+A^{0.5}_3)$ & $20A^{0.5}_1$ & $%
5(-11A^{0.5}_1+2A^{0.5}_3)$ & $-25A^{0.5}_1$ \\ \hline
$C_{3S1}^{0.5+}$ & $-20(6A^{0.5}_2+A^{0.5}_4)$ & $60A^{0.5}_2$ & $%
-25(2A^{0.5}_2-A^{0.5}_4)$ & $-29A^{0.5}_2$ \\ \hline
$C_{2S1}^{0.5+}$ & $-60(A^{0.5}_1+A^{0.5}_3)$ & $20(4A^{0.5}_1+3A^{0.5}_3)$ & $%
-5(13A^{0.5}_1-14A^{0.5}_3)$ & $2(-6A^{0.5}_1+7A^{0.5}_3)$ \\ \hline
$C_{1S1}^{0.5+}$ & $-60A^{0.5}_2$ & $20(9A^{0.5}_2+A^{0.5}_4)$ & $75A^{0.5}_2$
& $81A^{0.5}_2+17A^{0.5}_4$ \\ \hline
$C_{0S1}^{0.5+}$ & $-20A^{0.5}_1$ & $60(A^{0.5}_1+A^{0.5}_3)$ & $7A^{0.5}_1$ &
$21A^{0.5}_1+50A^{0.5}_3$ \\
\hline\hline
$S_{6S3}^{0.5+}$ & $A^{0.5}_1$ & $0$ & $4A^{0.5}_1$ & $0$ \\ \hline
$S_{5S3}^{0.5+}$ & $-A^{0.5}_2$ & $0$ & $12A^{0.5}_2$ & $0$ \\ \hline
$S_{4S3}^{0.5+}$ & $-11A^{0.5}_1+2A^{0.5}_3$ & $11A^{0.5}_1$ & $%
12(A^{0.5}_1+A^{0.5}_3)$ & $-4A^{0.5}_1$ \\ \hline
$S_{3S3}^{0.5+}$ & $-5(2A^{0.5}_2-A^{0.5}_4)$ & $23A^{0.5}_2$ & $%
4(6A^{0.5}_2+A^{0.5}_4)$ & $-12A^{0.5}_2$ \\ \hline
$S_{2S3}^{0.5+}$ & $(-13A^{0.5}_1+14A^{0.5}_3)$ & $2(5A^{0.5}_1+11A^{0.5}_3)$
& $12(A^{0.5}_1+A^{0.5}_3)$ & $-4(2A^{0.5}_1+3A^{0.5}_3)$ \\ \hline
$S_{1S3}^{0.5+}$ & $15A^{0.5}_2$ & $A^{0.5}_2+13A^{0.5}_4$ & $12A^{0.5}_2$ & $%
-4(3A^{0.5}_2+A^{0.5}_4)$ \\
\hline\hline
$S_{6S1}^{0.5+}$ & $5A^{0.5}_1$ & $0$ & $20A^{0.5}_1$ & $0$ \\ \hline
$S_{5S1}^{0.5+}$ & $-5A^{0.5}_2$ & $0$ & $60A^{0.5}_2$ & $0$ \\ \hline
$S_{4S1}^{0.5+}$ & $-5(11A^{0.5}_1-2A^{0.5}_3)$ & $-25A^{0.5}_1$ & $%
60(A^{0.5}_1+A^{0.5}_3)$ & $-20A^{0.5}_1$ \\ \hline
$S_{3S1}^{0.5+}$ & $-25(2A^{0.5}_2-A^{0.5}_4)$ & $-29A^{0.5}_2$ & $%
20(6A^{0.5}_2+A^{0.5}_4)$ & $-60A^{0.5}_2$ \\ \hline
$S_{2S1}^{0.5+}$ & $5(-13A^{0.5}_1+14A^{0.5}_3)$ & $2(-23A^{0.5}_1+7A^{0.5}_3)$
& $60(A^{0.5}_1+A^{0.5}_3)$ & $-20(2A^{0.5}_1+3A^{0.5}_3)$ \\ \hline
$S_{1S1}^{0.5+}$ & $75A^{0.5}_2$ & $-21A^{0.5}_2+17A^{0.5}_4$ & $60A^{0.5}_2$
& $-20(3A^{0.5}_2+A^{0.5}_4)$ \\ \hline
\end{tabular}%
\end{center}
\caption{This table contains the relevant $a^{C_F}_l$, $a^{S_F}_l$,
$b^{C_F}_l$, and $b^{S_F}_l$ coefficients for evaluating the
explicit form of $h^{0.5+}$ in Eq.(\protect\ref{h05open}).}
\label{coeff:open05plus}
\end{table}

\begin{table}[!ht]
\begin{center}
\begin{tabular}{|c||c|c|c|c|}
\hline Harmonic & $\cos{[3\Upsilon_0]}$ & $\cos{[\Upsilon_0]}$ &
$\sin{[3\Upsilon_0]}$ & $\sin{[\Upsilon_0]}$ \\ \hline\hline
$C_{6S2}^{0.5\times}$ & $-4A^{0.5}_1$ & $0$ & $-16A^{0.5}_1$ & $0$
\\ \hline $C_{5S2}^{0.5\times}$ & $4A^{0.5}_2$ & $0$ &
$-48A^{0.5}_2$ & $0$ \\ \hline
$C_{4S2}^{0.5\times}$ & $4(11A^{0.5}_1-2A^{0.5}_3)$ & $-4A^{0.5}_1$ & $%
-48(A^{0.5}_1+A^{0.5}_3)$ & $16A^{0.5}_1$ \\ \hline
$C_{3S2}^{0.5\times}$ & $20(2A^{0.5}_2-A^{0.5}_4)$ & $-20A^{0.5}_2$ & $%
-16(6A^{0.5}_2+A^{0.5}_4)$ & $48A^{0.5}_2$ \\ \hline
$C_{2S2}^{0.5\times}$ & $4(13A^{0.5}_1-14A^{0.5}_3)$ & $%
-8(6A^{0.5}_1+5A^{0.5}_3)$ & $-48(A^{0.5}_1+A^{0.5}_3)$ & $%
16(4A^{0.5}_1+3A^{0.5}_3)$ \\ \hline
$C_{1S2}^{0.5\times}$ & $-60A^{0.5}_2$ & $-4(39A^{0.5}_2+7A^{0.5}_4)$ & $%
-48A^{0.5}_2$ & $16(9A^{0.5}_2+A^{0.5}_4)$ \\ \hline
$C_{0S2}^{0.5\times}$ & $-28A^{0.5}_1$ & $-4(3A^{0.5}_1+22A^{0.5}_3)$ & $%
-16A^{0.5}_1$ & $48(A^{0.5}_1+A^{0.5}_3)$ \\ \hline
$S_{6S2}^{0.5\times}$ & $-16A^{0.5}_1$ & $0$ & $4A^{0.5}_1$ & $0$ \\ \hline
$S_{5S2}^{0.5\times}$ & $-48A^{0.5}_2$ & $0$ & $-4A^{0.5}_2$ & $0$ \\ \hline
$S_{4S2}^{0.5\times}$ & $-48(A^{0.5}_1+A^{0.5}_3)$ & $16A^{0.5}_1$ & $%
4(-11A^{0.5}_1+2A^{0.5}_3)$ & $4A^{0.5}_1$ \\ \hline
$S_{3S2}^{0.5\times}$ & $-16(6A^{0.5}_2+A^{0.5}_4)$ & $48A^{0.5}_2$ & $%
-20(2A^{0.5}_2-A^{0.5}_4)$ & $20A^{0.5}_2$ \\ \hline
$S_{2S2}^{0.5\times}$ & $-48(A^{0.5}_1+A^{0.5}_3)$ & $%
16(2A^{0.5}_1+3A^{0.5}_3) $ & $4(-13A^{0.5}_1+14A^{0.5}_3)$ & $%
-8(A^{0.5}_1-5A^{0.5}_3)$ \\ \hline
$S_{1S2}^{0.5\times}$ & $-48A^{0.5}_2$ & $16(3A^{0.5}_2+A^{0.5}_4)$ & $%
60A^{0.5}_2$ & $4(-3A^{0.5}_2+7A^{0.5}_4)$ \\ \hline
\end{tabular}%
\end{center}
\caption{This table contains the relevant $a^{C_F}_l$, $a^{S_F}_l$,
$b^{C_F}_l$, and $b^{S_F}_l$ coefficients for evaluating the
explicit form of $h^{0.5\times}$ in Eq.(\protect\ref{h05open}).}
\label{coeff:open05cross}
\end{table}

\begin{table}[!ht]
\begin{center}
\begin{tabular}{|c||c|c|c|}
\hline Harmonic & $\cos{[4\Upsilon_0]}$ & $\cos{[2\Upsilon_0]}$ &
$1$
\\ \hline\hline $C^{1+}_{7C4}$ & $15KA^1_2$ & $0$ & $0$ \\ \hline
$C^{1+}_{6C4}$ & $35KA^1_3$ & $0$ & $0$ \\ \hline $C^{1+}_{5C4}$ &
$81KA^1_2+452KA^1_4$ & $-24KA^1_2$ & $0$ \\ \hline $C^{1+}_{4C4}$ &
$142KA^1_3+8KA^1_5$ & $-40KA^1_3$ & $0$ \\ \hline $C^{1+}_{3C4}$ &
$189KA^1_2+1076KA^1_4$ & $-144KA^1_2-256KA^1_4$ & $30KA^1_2$
\\ \hline
$C^{1+}_{2C4}$ & $215KA^1_3$ & $-152KA^1_3+2KA^1_5$ & $30KA^1_3$ \\ \hline
$C^{1+}_{1C4}$ & $315KA^1_2$ & $-504KA^1_2+224KA^1_4$ & $234KA^1_2-24KA^1_4$
\\ \hline
$C^{1+}_{0C4}$ & $KA^1_1$ & $-4KA^1_1+32KA^1_3$ & $3KA^1_1-6KA^1_3$ \\
\hline\hline
$C^{1+}_{7C2}$ & $60KA^1_2$ & $0$ & $0$ \\ \hline
$C^{1+}_{6C2}$ & $140KA^1_3$ & $0$ & $0$ \\ \hline
$C^{1+}_{5C2}$ & $324KA^1_2+1808KA^1_4$ & $96(2-9\eta)A^1_2$ & $0$ \\ \hline
$C^{1+}_{4C2}$ & $568KA^1_3+32KA^1_5$ & $8(4-123\eta)A^1_3$ & $0$ \\ \hline
$C^{1+}_{3C2}$ & $756KA^1_2+4304KA^1_4$ & $96(5-14\eta)A^1_2-64(13+114%
\eta)A^1_4$ & $24(29-15\eta)A^1_2$ \\ \hline
$C^{1+}_{2C2}$ & $860KA^1_3$ & $16(34-123\eta)A^1_3-8(2+9\eta)A^1_5$ & $%
120(1+3\eta)A^1_3$ \\ \hline
$C^{1+}_{1C2}$ & $1260KA^1_2$ & $96(33-61\eta)A^1_2+64(29-120\eta)A^1_4$ & $%
24(235\eta-33)A^1_2-96(37-35\eta)A^1_4$ \\ \hline
$C^{1+}_{0C2}$ & $4KA^1_1$ & $8KA^1_1+8(100-171\eta)A^1_3$ & $%
-12KA^1_1-24(13-49\eta)A^1_3-12(3-\eta)A^1_5$ \\ \hline\hline
$C^{1+}_{70}$ & $-75KA^1_2$ & $0$ & $0$ \\ \hline
$C^{1+}_{60}$ & $-175KA^1_3$ & $0$ & $0$ \\ \hline
$C^{1+}_{50}$ & $405KA^1_2+2260KA^1_4$ & $24(1-39\eta)A^1_2$ & $0$ \\ \hline
$C^{1+}_{40}$ & $-710KA^1_3-40KA^1_5$ & $-8(103+24\eta)A^1_3$ & $0$ \\ \hline
$C^{1+}_{30}$ & $-945KA^1_2-5380KA^1_4$ & $48(39\eta-11)A^1_2-64(167-42%
\eta)A^1_4$ & $6(75\eta-121)A^1_2$ \\ \hline
$C^{1+}_{20}$ & $-1075KA^1_3$ & $8(69\eta-65)A^1_3-2(73-39\eta)A^1_5$ & $%
-30(5+9\eta)A^1_3$ \\ \hline
$C^{1+}_{10}$ & $-1575KA^1_2$ & $72(91-121\eta)A^1_2-32(145-237\eta)A^1_4$ &
$6(93-823\eta)A^1_2+24(149-143\eta)A^1_4$ \\ \hline
$C^{1+}_{00}$ & $-5KA^1_1$ & $32KA^1_1+8(128+3\eta)A^1_3$ & $%
9KA^1_1+6(53-199\eta)A^1_3+12(3-\eta)A^1_5$ \\ \hline
\end{tabular}%
\end{center}
\caption{This table contains all relevant coefficients for evaluating every $C_F$
in Eq.(\protect\ref{h1open}) for $h^{1+}$.}
\label{coeff:open1plusC}
\end{table}

\begin{table}[!ht]
\begin{center}
\begin{tabular}{|c||c|c|}
\hline Harmonic & $\sin{[4\Upsilon_0]}$ & $\sin{[2\Upsilon_0]}$ \\
\hline\hline $S^{1+}_{7C4}$ & $-15KA^1_2$ & $0$ \\ \hline
$S^{1+}_{6C4}$ & $-35KA^1_3$ & $0$ \\ \hline $S^{1+}_{5C4}$ &
$-81KA^1_2-452KA^1_4$ & $24KA^1_2$ \\ \hline $S^{1+}_{4C4}$ &
$-142KA^1_3-8KA^1_5$ & $40KA^1_3$ \\ \hline $S^{1+}_{3C4}$ &
$-189KA^1_2-1076KA^1_4$ & $144KA^1_2+256KA^1_4$ \\ \hline
$S^{1+}_{2C4}$ & $-215KA^1_3$ & $152KA^1_3-2KA^1_5$ \\ \hline
$S^{1+}_{1C4}$ & $-315KA^1_2$ & $504KA^1_2-224KA^1_4$ \\
\hline\hline $S^{1+}_{7C2}$ & $-60KA^1_2$ & $0$ \\ \hline
$S^{1+}_{6C2}$ & $-140KA^1_3$ & $0$ \\ \hline $S^{1+}_{5C2}$ &
$-324KA^1_2-1808KA^1_4$ & $96(9\eta-2)A^1_2$ \\ \hline
$S^{1+}_{4C2}$ & $-568KA^1_3-32KA^1_5$ & $8(123\eta-4)A^1_3$ \\
\hline
$S^{1+}_{3C2}$ & $-756KA^1_2-4304KA^1_4$ & $96(14\eta-5)A^1_2+64(13+114%
\eta)A^1_4$ \\ \hline
$S^{1+}_{2C2}$ & $-860KA^1_3$ & $16(123\eta-34)A^1_3+8(2+9\eta)A^1_5$ \\
\hline
$S^{1+}_{1C2}$ & $-1260KA^1_2$ & $32(87\eta-33)A^1_2-64(29-120\eta)A^1_4)$
\\ \hline\hline
$S^{1+}_{70}$ & $75KA^1_2$ & $0$ \\ \hline
$S^{1+}_{60}$ & $175KA^1_3$ & $0$ \\ \hline
$S^{1+}_{50}$ & $-405KA^1_2-2260KA^1_4$ & $24(39\eta-1)A^1_2$ \\ \hline
$S^{1+}_{40}$ & $710KA^1_3+40KA^1_5$ & $8(103+24\eta)A^1_3$ \\ \hline
$S^{1+}_{30}$ & $945KA^1_2+5380KA^1_4$ & $48(11-39\eta)A^1_2+64(167-42%
\eta)A^1_4$ \\ \hline
$S^{1+}_{20}$ & $1075KA^1_3$ & $8(65-69\eta)A^1_3+2(73-39\eta)A^1_5$ \\
\hline
$S^{1+}_{10}$ & $1575KA^1_2$ & $24(219\eta-89)A^1_2+(145-237\eta)32A^1_4$ \\
\hline
\end{tabular}%
\end{center}
\caption{This table contains all relevant coefficients for evaluating every $S_F$
in Eq.(\protect\ref{h1open}) for $h^{1+}$.}
\label{coeff:open1plusS}
\end{table}

\begin{table}[!ht]
\begin{center}
\begin{tabular}{|c||c|c|}
\hline Harmonic & $\sin{[4\Upsilon_0]}$ & $\sin{[2\Upsilon_0]}$ \\
\hline\hline $C^{1\times}_{7C3}$ & $60KA^1_2$ & $0$ \\ \hline
$C^{1\times}_{6C3}$ & $140KA^1_3$ & $0$ \\ \hline
$C^{1\times}_{5C3}$ & $324KA^1_2+1808KA^1_4$ & $24KA^1_2$ \\ \hline
$C^{1\times}_{4C3}$ & $568KA^1_3+32KA^1_5$ & $40KA^1_3$ \\ \hline
$C^{1\times}_{3C3}$ & $756KA^1_2+4304KA^1_4$ & $-24KA^1_2+544KA^1_4$ \\
\hline
$C^{1\times}_{2C3}$ & $860KA^1_3$ & $-16KA^1_3+16KA^1_5$ \\ \hline
$C^{1\times}_{1C3}$ & $1260KA^1_2$ & $-576KA^1_2+1696KA^1_4$ \\ \hline
$C^{1\times}_{0C3}$ & $4KA^1_1$ & $-8KA^1_1+232KA^1_3$ \\ \hline\hline
$C^{1\times}_{7C1}$ & $-60KA^1_2$ & $0$ \\ \hline
$C^{1\times}_{6C1}$ & $-140KA^1_3$ & $0$ \\ \hline
$C^{1\times}_{5C1}$ & $-324KA^1_2-1808KA^1_4$ & $24(7-69\eta)A^1_2$ \\ \hline
$C^{1\times}_{4C1}$ & $-568KA^1_3-32KA^1_5$ & $-8(109+117\eta)A^1_3$ \\
\hline
$C^{1\times}_{3C1}$ & $-756KA^1_2-4304KA^1_4$ & $24(37\eta-7)A^1_2-32(385+69%
\eta)A^1_4$ \\ \hline
$C^{1\times}_{2C1}$ & $-860KA^1_3$ & $-112(1+9\eta)A^1_3-16(11-3\eta)A^1_5$
\\ \hline
$C^{1\times}_{1C1}$ & $-1260KA^1_2$ & $192(51-77\eta)A^1_2-32(133-135%
\eta)A^1_4$ \\ \hline
$C^{1\times}_{0C1}$ & $-4KA^1_1$ & $40KA^1_1+8(203-93\eta)A^1_3$ \\ \hline
\end{tabular}%
\end{center}
\caption{This table contains all relevant coefficients for evaluating every $C_F$
in Eq.(\protect\ref{h1open}) for $h^{1\times}$.}
\label{coeff:open1crossC}
\end{table}

\begin{table}[!ht]
\begin{center}
\begin{tabular}{|c||c|c|c|}
\hline Harmonic & $\cos{[4\Upsilon_0]}$ & $\cos{[2\Upsilon_0]}$ &
$1$
\\ \hline\hline $S^{1\times}_{7C3}$ & $60KA^1_2$ & $0$ & $0$ \\
\hline $S^{1\times}_{6C3}$ & $140KA^1_3$ & $0$ & $0$ \\ \hline
$S^{1\times}_{5C3}$ & $324KA^1_2+1808KA^1_4$ & $24KA^1_2$ & $0$ \\
\hline $S^{1\times}_{4C3}$ & $568KA^1_3+32KA^1_5$ & $40KA^1_3$ & $0$
\\ \hline
$S^{1\times}_{3C3}$ & $756KA^1_2+4304KA^1_4$ & $-24KA^1_2+544KA^1_4$ & $%
-48KA^1_2$ \\ \hline
$S^{1\times}_{2C3}$ & $860KA^1_3$ & $-16KA^1_3+16KA^1_5$ & $-48KA^1_3$ \\
\hline
$S^{1\times}_{1C3}$ & $1260KA^1_2$ & $-816KA^1_2+1696KA^1_4$ & $%
-48KA^1_2-192KA^1_4$ \\ \hline\hline
$S^{1\times}_{7C1}$ & $-60KA^1_2$ & $0$ & $0$ \\ \hline
$S^{1\times}_{6C1}$ & $-140KA^1_3$ & $0$ & $0$ \\ \hline
$S^{1\times}_{5C1}$ & $-324KA^1_2-1808KA^1_4$ & $24(7-69\eta)A^1_2$ & $0$ \\
\hline
$S^{1\times}_{4C1}$ & $-568KA^1_3-32KA^1_5$ & $-8(109+117\eta)A^1_3$ & $0$
\\ \hline
$S^{1\times}_{3C1}$ & $-756KA^1_2-4304KA^1_4$ & $24(37\eta-7)A^1_2-32(385+69%
\eta)A^1_4$ & $48KA^1_2$ \\ \hline
$S^{1\times}_{2C1}$ & $-860KA^1_3$ & $-112(1+9\eta)A^1_3-16(11-3\eta)A^1_5$
& $48KA^1_3$ \\ \hline
$S^{1\times}_{1C1}$ & $1260KA^1_2$ & $48(73-187\eta)A^1_2-32(133-135%
\eta)A^1_4$ & $48KA^1_2+192KA^1_4$ \\ \hline
\end{tabular}%
\end{center}
\caption{This table contains all relevant coefficients for evaluating every $S_F$
in Eq.(\protect\ref{h1open}) for $h^{1\times}$.}
\label{coeff:open1crossS}
\end{table}

\begin{table}[!ht]
\begin{center}
\begin{tabular}{|c||c|c|c|}
\hline Harmonic & $\cos{[2\Upsilon_0]}$ & $\sin{[2\Upsilon_0]}$ &
$1$
\\ \hline\hline $C^{N+}_{4C2}$ & $-BA^N_5$ & $0$ & $0$ \\ \hline
$C^{N+}_{3C2}$ & $-(31+13\eta)A^N_{10}-4(10-9\eta)A^N_{11}-2(14-17%
\eta)A^N_{12}$ & $0$ & $2(2-\eta)A^N_4$ \\ \hline
$C^{N+}_{2C2}$ & $2(-3+5\eta)A^N_5-10KA^N_9$ & $-4BA^N_5$ & $%
4(5-4\eta)A^N_5+28(1-\eta)A^N_9$ \\ \hline
$C^{N+}_{1C2}$ & $(89\eta-83)A^N_{10}-(48-55\eta)A^N_{11}-2(22-31%
\eta)A^N_{12}$ & $-2BA^N_4$ & $2(43-49\eta)A^N_{10}+4(20-17%
\eta)A^N_{11}+4(28-25\eta)A^N_{12}$ \\ \hline
$C^{N+}_{0C2}$ & $-12KA^1_1-(28-15\eta)A^N_5-(17-29\eta)A^N_9$ & $0$ & $%
12KA^1_1+3(9-5\eta)A^N_5+(33-31\eta)A^N_9$ \\ \hline\hline
$C^{N+}_{40}$ & $-3BA^N_5$ & $0$ & $0$ \\ \hline
$C^{N+}_{30}$ & $-3(31+13\eta)A^N_{10}-12(10-9\eta)A^N_{11}-6(14-17%
\eta)A^N_{12}$ & $0$ & $2(\eta-2)A^N_4$ \\ \hline
$C^{N+}_{20}$ & $6(5\eta-3)A^N_5-30KA^N_9$ & $-12BA^N_5$ & $%
4(4\eta-5)A^N_5-28(1-\eta)A^N_9$ \\ \hline
$C^{N+}_{10}$ & $3(89\eta-83)A^N_{10}-3(48-55\eta)A^N_{11}-6(22-31%
\eta)A^N_{12}$ & $-6BA^N_4$ & $2(49\eta-43)A^N_{10}-4(20-17%
\eta)A^N_{11}-4(28-25\eta)A^N_{12}$ \\ \hline
$C^{N+}_{00}$ & $-36KA^1_1-3(28-15\eta)A^N_5-3(17-29\eta)A^N_9$ & $0$ & $%
-12KA^1_1-3(9-5\eta)A^N_5-(33-31\eta)A^N_9$ \\ \hline
\end{tabular}%
\end{center}
\caption{1PN coefficients for $h^{N+}$ in Eq.(\protect\ref%
{hNopen}) from the 1PN description of the motion. This table
contains only the 1PN corrections to the coefficients needed to
evaluate the $C^N_F$ quantities.}
\label{coeff:openPNplusC}
\end{table}

\begin{table}[!ht]
\begin{center}
\begin{tabular}{|c||c|c|c|}
\hline Harmonic & $\cos{[2\Upsilon_0]}$ & $\sin{[2\Upsilon_0]}$ &
$1$
\\ \hline\hline $S^{N+}_{4C2}$ & $0$ & $BA^N_5$ & $0$ \\ \hline
$S^{N+}_{3C2}$ & $0$ & $(31+13\eta)A^N_{10}+4(10-9\eta)A^N_{11}+2(14-17%
\eta)A^N_{12}$ & $0$ \\ \hline
$S^{N+}_{2C2}$ & $-4BA^N_5$ & $2(3-5\eta)A^N_5+10KA^N_9$ & $0$ \\ \hline
$S^{N+}_{1C2}$ & $-2BA^N_4$ & $(83-89\eta)A^N_{10}+(48-55%
\eta)A^N_{11}+2(22-31\eta)A^N_{12}$ & $-2BA^N_4$ \\ \hline\hline
$S^{N+}_{40}$ & $0$ & $3BA^N_5$ & $0$ \\ \hline
$S^{N+}_{30}$ & $0$ & $3(31+13\eta)A^N_{10}+12(10-9\eta)A^N_{11}+6(14-17%
\eta)A^N_{12}$ & $0$ \\ \hline
$S^{N+}_{20}$ & $-12BA^N_5$ & $6(3-5\eta)A^N_5+30KA^N_9$ & $0$ \\ \hline
$S^{N+}_{10}$ & $-6BA^N_4$ & $3(83-89\eta)A^N_{10}+3(48-55%
\eta)A^N_{11}+6(22-31\eta)A^N_{12}$ & $6BA^N_4$ \\ \hline
\end{tabular}%
\end{center}
\caption{1PN coefficients for $h^{N+}$ in Eq.(\protect\ref%
{hNopen}) from the 1PN description of the motion. This table
contains only the 1PN corrections to the coefficients needed to
evaluate the $S^N_F$ quantities.}
\label{coeff:openPNplusS}
\end{table}

\begin{table}[!ht]
\begin{center}
\begin{tabular}{|c||c|c|}
\hline Harmonic & $\cos{[2\Upsilon_0]}$ & $\sin{[2\Upsilon_0]}$ \\
\hline\hline $C^{N\times}_{4}$ & $0$ & $-4BA^N_5$ \\ \hline
$C^{N\times}_{3}$ & $0$ & $-4(31+13\eta)A^N_{10}-16(10-9%
\eta)A^N_{11}-8(14-17\eta)A^N_{12}$ \\ \hline
$C^{N\times}_{2}$ & $16BA^N_5$ & $8(5\eta-3)A^N_5-40KA^N_9$ \\ \hline
$C^{N\times}_{1}$ & $8BA^N_4$ & $4(89\eta-83)A^N_{10}-4(48-55%
\eta)A^N_{11}-8(22-31\eta)A^N_{12}$ \\ \hline
$C^{N\times}_{0}$ & $0$ & $-16KA^1_1-12(10-9\eta)A^N_5-4(17-23\eta)A^N_9$ \\
\hline\hline
$S^{N\times}_{4}$ & $-4BA^N_5$ & $0$ \\ \hline
$S^{N\times}_{3}$ & $-4(31+13\eta)A^N_{10}-16(10-9\eta)A^N_{11}-8(14-17%
\eta)A^N_{12}$ & $0$ \\ \hline
$S^{N\times}_{2}$ & $8(5\eta-3)A^N_5-40KA^N_9$ & $-16BA^N_5$ \\ \hline
$S^{N\times}_{1}$ & $4(89\eta-83)A^N_{10}-4(48-55\eta)A^N_{11}-8(22-31%
\eta)A^N_{12}$ & $-8BA^N_4$ \\ \hline
\end{tabular}%
\end{center}
\caption{1PN coefficients for $h^{N\times}$ in Eq.(\protect\ref%
{hNopen}) from the 1PN description of the motion.}
\label{coeff:openPNcross}
\end{table}

For open orbits there are no difficulties when evaluating series
expansion, in the final formulae there are two different types of
terms: one contains higher harmonics of the true anomaly parameter,
but there are also terms $\sim \chi\cos k\chi / \chi\sin k\chi$. In
Table \ref{coeff:openPNchi} the coefficients $C_{k\chi}$ and
$S_{k\chi}$ of these terms are presented.
\begin{table}[!ht]
\begin{center}
\begin{tabular}{|c||c|c|c|c||c|c|}
\hline State & \multicolumn{4}{|c||}{$+$} &
\multicolumn{2}{|c|}{$\times$} \\ \hline $\gamma$-harmonic &
\multicolumn{2}{|c|}{$\cos{[2\gamma]}$} & \multicolumn{2}{|c||}{$1$}
 & \multicolumn{2}{|c|}{$\cos{[\gamma]}$} \\ \hline
$\Upsilon_0$-harmonic & $\cos{[2\Upsilon_0]}$ & $\sin{[2\Upsilon_0]}$ & $\cos{[2\Upsilon_0]}$
& $\sin{[2\Upsilon_0]}$ & $\cos{[2\Upsilon_0]}$ & $\sin{[2\Upsilon_0]}$ \\
\hline\hline
$C_{3\chi}$ & $0$ & $3A^N_6$ & $0$ & $9A^N_6$ & $-12A^N_6$ & $0$ \\ \hline
$C_{2\chi}$ & $0$ & $48A^N_9$ & $0$ & $144A^N_9$ & $-192A^N_9$ & $0$ \\
\hline
$C_{1\chi}$ & $0$ & $15A^N_6$ & $0$ & $45A^N_6$ & $-60A^N_6$ & $0$ \\ \hline
$C_{0\chi}$ & $0$ & $24A^N_5$ & $0$ & $72A^N_5$ & $-96A^N_5$ & $0$ \\ \hline
$S_{3\chi}$ & $3A^N_6$ & $0$ & $9A^N_6$ & $0$ & $0$ & $12A^N_6$ \\ \hline
$S_{2\chi}$ & $48A^N_9$ & $0$ & $144A^N_9$ & $0$ & $0$ & $192A^N_9$ \\ \hline
$S_{1\chi}$ & $15A^N_6$ & $0$ & $45A^N_6$ & $0$ & $0$ & $60A^N_6$ \\ \hline
\end{tabular}%
\end{center}
\caption{This table contains the 1PN coefficients for both
polarization states of Eq.(\protect\ref{hNopen}) from the 1PN
description of the motion which are multiplied by $\protect\chi$.
The column under the sign "$+$" contains the coefficients for
$h^{N+}$, and under the sign "$\times$" the coefficients for
$h^{N\times}$ can be found.} \label{coeff:openPNchi}
\end{table}

\end{document}